\documentclass[11pt]{article}

\usepackage{mathtools}
\usepackage{amssymb}
\usepackage{dsfont}
\usepackage{slashed}
\usepackage{fullpage}
\usepackage[colorlinks=true,linkcolor=blue,citecolor=magenta,linktocpage=true]{hyperref}
\usepackage{titlesec}
\titleformat*{\section}{\normalsize\bfseries}
\titleformat*{\subsection}{\normalsize\bfseries}
\titleformat*{\subsubsection}{\normalsize\bfseries}
\usepackage{cite}
\usepackage{cancel}

\DeclareMathAlphabet{\bbvar}{U}{BOONDOX-ds}{m}{n}

\makeatletter
\renewcommand{\@dotsep}{10000}
\makeatother

\newcommand{\cB}{{\mathcal B}}

\newcommand{\cK}{{\mathcal K}}
\newcommand{\cL}{{\mathcal L}}
\newcommand{\cH}{{\mathcal H}}

\newcommand{\cR}{{\mathcal R}}
\newcommand{\cO}{{\mathcal O}}
\newcommand{\cP}{{\mathcal P}}
\newcommand{\cQ}{{\mathcal Q}}

\newcommand{\cD}{{\mathcal D}}

\newcommand{\be}{\begin{equation}}
\newcommand{\ee}{\end{equation}}
\newcommand{\beq}{\begin{eqnarray}}
\newcommand{\eeq}{\end{eqnarray}}
\newcommand{\bes}{\begin{eqnarray}}
\newcommand{\ees}{\end{eqnarray}}

\newcommand{\mat} [2] {\left ( \begin{array}{#1}#2\end{array} \right ) }

\def\rd{\textrm{d}}

\def\beq{\begin{eqnarray}}
\def\eeq{\end{eqnarray}}
\def\be{\begin{equation}}
\def\ee{\end{equation}}

\def\r2A{z}

\usepackage[toc,page]{appendix}
\usepackage[usenames,dvipsnames]{xcolor}

\def\nn{\nonumber}

\numberwithin{equation}{section}

\begin{document}

\title{\Large{\textbf{\sffamily Jet launching from the Kerr black hole magnetosphere: \\ An electrogeodesic approach }}}
\author{\sffamily Jibril Ben Achour\;$^{1,2}$, Ileyk El Mellah\;$^{4}$, Eric Gourgoulhon\;$^{5,6}$}
\date{\small{\textit{$^{1}$ Arnold Sommerfeld Center for Theoretical Physics, Munich, Germany, \\
$^{2}$Univ Lyon, CNRS, ENS de Lyon, Laboratoire de Physique (LPENSL), Lyon, France\\
$^{4}$ Departament de Física, EEBE, Universitat Politècnica de Catalunya, c/Eduard Maristany 16, 08019 Barcelona, Spain\\
$^{5}$ LUX, Observatoire de Paris, Université PSL, CNRS, \\ Sorbonne Université, 92190 Meudon, France\\
$^{6}$ Laboratoire de Math\'ematiques de Bretagne Atlantique, CNRS UMR 6205, Universit\'e de Bretagne Occidentale, 6 avenue Victor Le Gorgeu, 29200 Brest, France
}}}

\maketitle

\begin{abstract}
The launch of relativistic jets of plasma on astrophysical to cosmological scales is observed in a variety of astrophysical sources, from active galactic nuclei to X-ray binaries. While these jets can be reproduced by general relativistic magneto-hydrodynamics (GRMHD) and particle-in-cells (GRPIC) simulations of the dynamical Kerr magnetosphere, the development of analytic models to describe the physics of the jets has remained limited. A key challenge is to analytically describe the individual trajectories of accelerated charged particles, which ultimately build up the jet and emit radiation. In this work, we provide a first simple but fully analytical model of jet launching from the Kerr magnetosphere based on the motion of charged particles. To that end, we use the integrability of electrogeodesic motion in the Kerr monopole magnetosphere to study the ejection of charged particles near the poles. This enables us to derive (i) a criterion for the rotation axis to constitute a stable latitudinal equilibrium position, thereby representing an idealized jet, (ii) the expression for the magnetic frame-dragging effect, and (iii) the condition for an asymptotic observer to measure blueshifted particles emanating from the black hole surroundings. Our study reveals that particles can be accelerated only in a specific region whose maximal radius depends on the spin and magnetization of the black hole. We further show that these results extend to the split-monopole Kerr solution by a redefinition of the angular momentum of the charged particles. Alongside these results, we provide a detailed review of the construction of test magnetospheres from (explicit and hidden) symmetries of the Kerr geometry and the condition for the separability of the electrogeodesic motion in a test magnetosphere, which serves as a basis for the model studied in this work.
\end{abstract}

\thispagestyle{empty}
\newpage
\setcounter{page}{1}

\hrule
\tableofcontents
\vspace{0.7cm}
\hrule


\newpage
\section{Introduction}

Relativistic jets of plasma are observed in a variety of astrophysical systems, from active galactic nuclei (AGN) containing a supermassive black hole (SMBH) in its centter to stellar mass black hole involved in X-ray binaries \cite{Doeleman:2012zc, Hada:2024icg}.  They can extend up to thousands of light years, therefore playing a major role in galaxy evolution by redistributing matter on large scales \cite{Oei:2024prz}. 
These jets are believed to be powered by highly magnetized spinning black holes which accrete plasma and accelerate particles at very high energy. Yet the precise mechanisms at play to produce and sustain these powerful jets, together with a detailed understanding of their structure, still remains to be understood \cite{Blandford:2018iot}. While catalogs of astrophysical systems powering jets have been monitored during the last decades \cite{Lister:2016ojc, Lister:2021wmw}, the newborn horizon-scale electromagnetic astronomy initiated by the Event Horizon Telescope \cite{EventHorizonTelescope:2019dse, EventHorizonTelescope:2022wok} now enables us to image relativistic jets of various systems with an unprecedented resolution, among which the M87 SMBH \cite{EventHorizonTelescope:2025uqi}. These observations coupled to a better characterization of the properties of the jet could provide a new window to further constrain the mass and spin of SMBHs \cite{Nokhrina:2019sxv}. 

To a large extent, our theoretical understanding of energy extraction from highly magnetized spinning black hole relies on the seminal Blandford-Znajek (BZ) model \cite{Blandford:1977ds}. The BZ solution describes a slowly rotating monopole magnetosphere surrounding the Kerr black hole, which is sourced by a dilute plasma. This Kerr magnetosphere successfully generates outgoing Poynting fluxes triggered by the toroidal components of the magnetic field. The BZ model relies on the force-free (FF) regime where the electromagnetic field satisfies $E \cdot B =0$ and $E^2 < B^2$ everywhere. The FF approximation ensures that the plasma density $\rho$ is sufficient to source the magnetosphere but small enough to keep it in a high magnetization regime, i.e. $\sigma \sim B^2/\rho \gg 1$. See \cite{Komissarov:2008yh, Ruiz:2012te, Toma:2014kva, Kinoshita:2017mio, Okamoto:2024fxa, Toma:2024tan} for more details on the BZ process. While the validity of the FF approximation and the asymptotic stability of the BZ solution casted some doubts in the early days \cite{Punsly:1989zz}, the progress in general relativistic (GR) magneto-hydrodynamics (MHD) simulations of dynamical black hole magnetospheres \cite{Koide:1999bj, Nishikawa:2004wp, McKinney:2004ka} allowed to address these questions and eventually numerically reproduce the final steady-state BZ solution and its Poynting winds \cite{Komissarov:2004qu, Komissarov:2005wj, Hawley:2005xs, Tchekhovskoy:2011zx}. See \cite{Gralla:2014yja} for a pedagogical review on the theoretical description of force-free magnetospheres.

Nevertheless, the BZ model and its associated GRMHD simulations suffer from several limitations to properly characterize the jet launching mechanism. First, the BZ solution is perturbative in the black hole spin and thus cannot account for rapidly spinning black holes. So far, no non-perturbative extension has been found \cite{Camilloni:2022kmx, Grignani:2018ntq, Pan:2015haa, Zhang:2014pla}. Second, and more importantly, the GRMHD simulations rely on a fluid description of the plasma which is inherently incomplete. In particular, it fails to track the motion of individual charged particles in the magnetosphere, which requires a kinetic approach. Switching from the fluid-based to a particle-based description of the jet is crucial to make progress. Moreover, the FF condition forbids the formation of pairs and electric gaps which are believed to play an important role in the jet dynamics. Even more importantly, the FF condition prevents charged particles to locally accelerate, which happens only when $E \cdot B \neq 0$, i.e. in the presence of non-ideal electric fields. The developments of GR/Particle-In-Cell (PIC) simulations have played a major role in addressing these difficulties \cite{Daniel:1997nu, Levinson:2018arx, Chen:2018khs, Hirotani:2020ubl, Yuan:2025war, Chernoglazov:2023ksi, Luepker:2025efw, Meringolo:2025bdu, Galishnikova:2025pfi, Chen:2025qio, Meringolo:2026qjf}. These early simulations \cite{Daniel:1997nu, Levinson:2018arx, Chen:2018khs}, based on the ZELTRON code, allow us to follow the individual particle trajectories of the plasma evolving in a dynamical magnetosphere where the force free condition can be violated. While they could also reproduce jet launching from a Kerr black hole immersed in an asymptotic uniform field \cite{Parfrey:2018dnc}, they also revealed new effects such as the build up of current sheet at the equator where magnetic reconnection appears to play a crucial role to eject energy from the black hole. See \cite{Chen:2019osy, Crinquand:2020ppq, Crinquand:2020reu, ElMellah:2021tjo, ElMellah:2023sun, Mehlhaff:2025mxi, Figueiredo:2025xbo} for further results along this line and \cite{Chen:2025qio, Meringolo:2026qjf} for alternatives available codes such as APERTURE, ENTITY and FPIC. Besides their qualitative results, these simulations underline the crucial need to properly characterize the motion of individual charged particles (as opposed to fluid) in the black hole magnetosphere to properly account of the complicated mechanisms at play in the accretion/ejection paradigm. 

The important progresses in the development of the GR/PIC simulations of the black hole magnetosphere dramatically contrast with the analytic investigations, which have remained scarce \cite{Gralla:2015wva, Mizuno:2025mog}. Indeed, the inherent complexity in solving the electrogeodesic motion around magnetized black holes has limited most studies to a mixture of analytical and numerical treatments \cite{Kolos:2015iva, Tursunov:2016dss, Kopacek:2018lgy, Kopacek:2020scv, Kopacek:2021lnq, Capitanio:2022epf, Khan:2023ttq, Kolos:2023oii, Rueda:2024xeb, Cherubini:2025lnc, Dyson:2023ujk}. This prevents us to obtain \textit{close analytic predictions} for the structure of the jet, its opening angle, the size of the acceleration and escape zones or the azimuthal velocity of charged particles in the different regions of the jet, which should play a key role to describe the emitted radiation. In contrast, the neutral geodesic motion in the Kerr geometry is well known to be fully integrable analytically \cite{Carter:1968rr}. The existence of a hidden symmetry, leading to the famous Carter constant, allows one to fully separate the radial and polar motion, which in turn, allows one to separate the time and azimuthal motion. In general, this two step process breaks down when a test-field magnetosphere is turned on, such that the electrogeodesic equations are no longer separable. This is typically the case of the well-known and extensively studied Wald solution \cite{Wald:1974np}, which describes an asymptotic uniform magnetic field around a Kerr black hole. However, the situation is not completely hopeless. 

Indeed, as initially shown in the seminal work of Carter \cite{Carter:1968rr}, the separability of the geodesic motion on the Kerr black hole extends to the electrogeodesic motion on the electrically and magnetically charged Kerr-Newman (KN) black hole, i.e. the electrovacuum solution to the Einstein-Maxwell equations
extending the Kerr solution and 
parametrized by four constants: the mass $M$, the spin parameter $a$, the electric charge $\cQ$ and the magnetic charge $\cP$. To our knowledge, the only systematic effort to extend Carter's analytic result and classify the electrogeodesic motion in the KN geometry was presented much later by Hackman and Xu in \cite{Hackmann:2013pva}. See \cite{Grunau:2010gd, Olivares:2011xb, Pugliese:2013zma, Das:2016opi} for related but more specific investigations. At first, it would appear natural to use these elegant results to study the jet formation in the KN magnetosphere. However, the presence of an electric charge $\cQ$ makes the model not so realistic, 
since the surrounding plasma would rapidly discharge the black hole. As for the magnetic charge, this limitation does not exist given that no elementary particles carrying magnetic charge are known to exist. Yet, the magnetic charge $\cP$ of the black hole is expected to be negligible compared to its mass, i.e. $\cP/M \ll 1$, so that gravity remains the dominant effect. This more realistic regime, i.e. $\cQ=0$ and $\cP/M \ll 1$, corresponds to the test field approximation of the magnetic KN geometry. Interestingly, taking this limit reduces the problem to the magnetically charged, vacuum magnetosphere on the Kerr black hole  which, although well known, has not received much attention in the literature. Yet, it appears to be the simplest analytic starting point to model a rotating black hole vacuum magnetosphere. At this level, a remarkable observation is that the separability of the electrogeodesic motion in the KN black hole remains valid in the test field approximation. This allows one to obtain closed form solutions to the equations of motion of charged particles. In this work, we use this path to analytically study the acceleration and the ejection of test charged particles around a magnetized Kerr black hole. Surprisingly, even in this simple vacuum magnetosphere, the exact electrogeodesic solutions reveal a rich behavior and allows one to identify a subset of trajectories that build up a jet of accelerated particles. Moreover, the interesting outcome is that this magnetically charged solution can be used to build the well known exact split-monopole solution (\ref{spli}) which does not carry any magnetic charge. As we shall see, the results derived here apply to both solutions. Let us summarize the new qualitative results for the physics of the jet emerging from this study:
\begin{itemize}
\item We present the analytic expression for the $4$-velocity $u^{\mu}$ [Eq.~(\ref{4velocity_Carter_tetrad}) below] and 4-acceleration $a^{\mu}$ [Eq.~(\ref{acc})] of charged particles in the whole black hole exterior, which allows us to contemplate several key properties of the system. Contrary to the azimuthal and polar acceleration $(a^{\theta}, a^{\varphi}$), the radial acceleration $a^r$ is shown to be proportional to both the magnetic charge $\cP$ (carried by the magnetosphere) and the black hole spin $a$, such that it vanishes in the non-rotating and/or non-magnetized case. This demonstrates the key role played by the presence of a spinning magnetically charged magnetosphere in the ejection mechanism. Second, the polar acceleration turns out to be maximal at the equator and vanishing at the pole. In contrast, the radial acceleration vanishes at the equator and becomes maximal at the poles, suggesting the build up of jet of charged particles preferentially around the poles. For negative charged particles, the acceleration is outwards and the particles are ejected if $\cP >0$.
\item To understand the structure of the ejection at the poles, and thus of the jet, we further analyze the stable and unstable equilibrium positions of the polar motion depending on the specific energy and angular momentum $(\varepsilon, \ell)$ of the charged particle. The presence of the magnetosphere dramatically changes the properties of the polar potential. The difference with the geodesic motion is best parametrized by a new parameter $\beta = - \kappa \cP /a$ where $\kappa$ is the charge-to-mass ratio of the particle.  First, particles can reach the Northern rotation axis $\theta=0$ only for $\ell =  a\beta$. Focusing on this family of electrogeodesics, we identify a new condition [Eq.~(\ref{theta_0_stable}) below] for the Northern rotation axis to be a stable latitudinal position, namely the energy of the particle must be either smaller or greater than a threshold depending on $\beta$. Instead, trajectories violating this condition admit a stable latitudinal position at some $\theta_{\ast} \in \; ] 0, \pi/2[$. 
\item We also study the fate of the frame-dragging effect in the Kerr monopole. Using the expression of the azimuthal angular velocity [Eq.~(\ref{MFD}) below], we show that the magnetic monopole introduces a new magnetic frame-dragging which completely dominates the gravitational frame-dragging. Indeed, we find that the former decays as $1/r^2$ while the later decays as $1/r^3$, showing that the new magnetic frame-dragging impacts charged particles on larger scales. Furthermore, in contrast to the gravitational frame-dragging which forces particles to rotate in a prograde trajectory, the magnetic counterpart forces particles to have prograde or retrograde motion depending on their charge, i.e. on $\kappa$. This magnetic frame dragging is expected to have interesting consequences on the precession of a misaligned jet of charged particles \cite{Cui:2023uyb}.
\item We further characterize the condition under which charged particles ejected from the magnetosphere reach an asymptotic observer with a higher energy. We evaluate the redshift experienced by the particle and find the condition for the particle to be blueshifted at large distance. Interestingly, this condition reveals the existence of a structure within the magnetosphere. It manifests through the existence of a radius which encodes the maximal extent of the region in which emitted particles escaping to infinity will be observed blueshifted. 
\end{itemize}
These different points, which rely on explicit analytic expressions, reveal a fine structure of the ejection mechanism of charged particles in the Kerr magnetosphere. It provides a first fully analytical study of the jet launching mechanism from the Kerr magnetosphere based on the exact solutions to the electrogeodesic motion. 

Since this investigation is made possible thanks to the separability of the electrogeodesic motion in this specific model of the Kerr magnetosphere, it is interesting to understand the fundamental reasons why this elegant integrability occurs. While this is well known for the Kerr-Newman geometry, it is interesting to revisit this question from the point of view of the test field approximation. Indeed, other exact solutions such as the Wald solution, which has been extensively studied, fail to have separable electrogeodesic motion. To make this work as self-contained as possible, we present, prior to our study of 
charged particle ejection, a review of (i) the construction of exact solutions of the Maxwell equations in the Kerr geometry using the Killing symmetries and (ii) 
a detailed derivation of the geometrical conditions for preserving the separability, i.e. the Carter constant, for the electrogeodesic motion in a given test field magnetosphere.  We also include a detailed derivation of the separable electrogeodesic equations in the Kerr magnetic monopole magnetosphere. This review allows us to present, besides the well known Wald construction, other results which have not received the same focus, such as the construction of exact vacuum solutions of Maxwell equations using Killing-Yano tensors known as the Penrose currents. We also use this opportunity to introduce new results, such as a non-vacuum solution of the Maxwell equation in the Kerr geometry built upon the two main Killing vectors [Eqs.~(\ref{newsol})--(\ref{newcurrent}) below]. In the end, this review provides the necessary technical tools to appreciate the fundamental reasons ensuring the separability of the electrogeodesic motion in the Kerr monopole magnetosphere, which descends from the properties of the Killing-Yano tensor of the Kerr geometry.

This article is organized as follows. In Section~\ref{secA}, we present a detailed review of the construction of exact test magnetosphere models from the explicit and hidden Killing symmetries, i.e. using the Killing vectors and the Killing-Yano tensors and apply them to the Kerr geometry. We present an alternative to the Wald construction in Section~\ref{A3}. Section~\ref{B} is devoted to the review on the separability of the electrogeodesic motion. We first review how the standard conditions for the geodesics are modified for the electrogeodesic case. Then, we derive the geometric condition for preserving the Carter constant. The detailed demonstration of the separability in the Kerr monopole magnetosphere is rederived in Section~\ref{B4}, together with the integral solutions for the electrogeodesic. In Section~\ref{C}, we apply this result to study the ejection of charged particles from the Kerr magnetosphere and the structure of the jet. Finally, Section~\ref{D} provides a discussion of our results, as well as of the open directions to explore.

\section{Kerr magnetospheres from symmetries}

\label{secA}

In this section, we review the key constructions allowing one to obtain exact vacuum solutions of Maxwell equations using the symmetries of the spacetime. 
Consider therefore the Maxwell equations in a curved spacetime; they take the form  
\begin{align}
\label{M1}
\nabla_{\nu} F^{\mu\nu} & = 4\pi J^{\mu} \\
\label{M2}
\nabla_{\mu} F_{\nu\alpha} + \nabla_{\nu} F_{\alpha\mu} + \nabla_{\alpha} F_{\mu\nu} & =0 ,
\end{align}
where the Faraday tensor $F_{\mu\nu}$ can be expressed in term of the gauge-potential $A =A_{\mu} \rd x^{\mu}$ as
\be
F_{\mu\nu} = \partial_{\mu} A_{\nu} - \partial_{\nu} A_{\mu} .
\ee
See \cite{Tsagas:2004kv} for a review of Maxwell theory in curved spacetime. 

In the following, we shall review how to construct specific solutions of these equations. The focus will be on the Kerr metric, which reads
\begin{align}
\rd s^2 & = - \left( 1 - \frac{2M r}{\Sigma}\right) \rd t^2 - \frac{4 a M r \sin^2{\theta}}{\Sigma} \, \rd t \, \rd \varphi + \frac{\Sigma}{\Delta} \rd r^2 + \Sigma \, \rd \theta^2 \nn \\
& \qquad \qquad + \left( r^2 + a^2 + \frac{2 a^2 M r \sin^2{\theta}}{\Sigma}\right) \sin^2{\theta} \, \rd \varphi^2
\end{align}
where the functions $\Sigma$ and $\Delta$ read
\be
\Sigma = r^2 + a^2 \cos^2{\theta} \qquad \Delta = r^2 - 2 M r + a^2 = (r- r_{+}) (r-r_{-})
\ee
with $r_{\pm}$ the positions of the outer and inner black hole horizons. 
It will also be useful to write its inverse form which reads
\begin{align}
g^{tt} & = - \frac{1}{\Delta} \left( r^2 +a^2 + \frac{2 a^2 M r \sin^2{\theta}}{\Sigma}\right) \qquad g^{\varphi\varphi} = \frac{1}{\Delta \sin^2{\theta}} \left( 1 - \frac{2M r}{\Sigma} \right) \\
g^{t\varphi} & = - \frac{2M a r }{\Sigma \Delta} \qquad g^{rr} = \frac{\Delta}{\Sigma} \qquad g^{\theta\theta} = \frac{1}{\Sigma}
\end{align}
The Kerr geometry is well known to have two commuting Killing vectors $\partial_t$ and $\partial_{\varphi}$. It also enjoys hidden symmetries generated by the presence of (conformal) Killing-Yano tensors. In the following, we shall review how both the  Killing vectors and the Killing-Yano tensors encode the solutions of the Maxwell equations. In particular, this will allow us to illustrate the role of the Killing tower structure of the Kerr geometry in relating different known solutions of the Maxwell equations. See \cite{Jezierski:2005cg, Krtous:2007xg, Penrose:1982wp, Jezierski:2015lwa, Kolar:2015cha} for more details on the Killing tower construction.

\subsection{Maxwell solutions from Killing vectors}

Consider a Killing vector $k^{\mu}$ on spacetime. By definition, it obeys  $\nabla_{(\mu} k_{\nu)} = 0$, which implies that it also satisfies the following equations:
\begin{align}
\Box k_{\mu} = - R_{\mu\nu} k^{\nu} \qquad \text{and} \qquad \nabla_{\mu} k^{\mu} =0 .
\end{align}
where $\Box = g^{\mu\nu} \nabla_{\mu} \nabla_{\nu}$ in term of the metric and the covariant derivative $\nabla$.
Writing down the source-free ($J^\mu=0$) Maxwell equation (\ref{M1}) in terms of the potential $A_{\mu}$, one finds 
\begin{align}
\Box A_{\mu} = R_{\mu\nu} A^{\nu} .
\end{align}
Therefore, upon working in the Lorentz gauge where $\nabla_{\mu} A^{\mu}  =0$, it is direct to see that any Killing vector of a metric obeying the vacuum Einstein equation $R_{\mu\nu}=0$
provides an exact analytic solution for the source-free Maxwell equation by setting $A_\mu = \alpha k_\mu$, where $\alpha$ is a constant. Hence, one obtains a very straightforward way to relate the profile of the Maxwell field to the explicit symmetries of the vacuum geometry. 

The Kerr metric being stationary and axisymmetric, it possesses two Killing vectors:
\be \label{Kerr_Killing_vectors}
\xi = \partial_t \qquad \text{and} \qquad  \chi = \partial_{\varphi} ,
\ee
which naturally give rise to a two parameters family of source-free Maxwell solutions of the form
\begin{align}
\label{SOL}
A^{(\alpha, \beta)}_{\mu}  = \alpha \xi_{\mu}  + \beta \chi_{\mu} .
\end{align}
Subsectors of this family of solutions have been first studied by Wald\footnote{See \cite{ Bicak:2006hs} for an extension where the magnetosphere is misaligned w.r.t the black hole spin.} in \cite{Wald:1974np}. In the following, we discuss the different branches of solutions and their properties. 

Before presenting the form of the electric and magnetic fields described by these solutions, it will reveal useful to introduce the Carter tetrad associated to the family of Carter observers.  
Each observer in this family is located at a fixed radial and polar coordinates $(r,\theta)$ and is rotating around the black hole at the 
angular velocity 
\be \label{Omega_Carter}
\Omega_{\rm C} = \frac{\rd\varphi}{\rd t} = \frac{a}{r^2 + a^2} . 
\ee
The 4-velocity of a Carter observer is a linear combination of the two Killing vectors: $e_0^\mu = (e_0)^t (\xi^\mu + \Omega_{\rm C} \chi^\mu)$ and the tetrad 
carried by that observer is the orthonormal vector frame 
$(e_\alpha)$ defined in terms of the Boyer-Lindquist coordinates by 
\begin{subequations}
\label{Carter_tetrad}
\begin{align}
    e_0 & = \frac{r^2 + a^2}{\sqrt{\Sigma \Delta}} \, \partial_t + \frac{a}{\sqrt{\Sigma \Delta}} \, \partial_\varphi \label{e_0_BL} \\
    e_1 & = \sqrt{\frac{\Delta}{\Sigma}} \, \partial_r \\
    e_2 & = \frac{1}{\sqrt{\Sigma}} \, \partial_\theta \\
    e_3 & = \frac{a\sin\theta}{\sqrt{\Sigma}} \, \partial_t + \frac{1}{\sqrt{\Sigma} \sin\theta} \, \partial_\varphi . 
\end{align}
\end{subequations}
The Carter observer has the distinctive feature that both the ingoing and outgoing principal null geodesics of the Kerr metric are purely radial with respect to him, i.e. their tangent vectors
have no component along $e_2$ and $e_3$. This explains why the Carter observer is well adapted to the description of the Kerr spacetime (outside the black hole region). 
Moreover, in the asymptotic region, i.e. for $r\to +\infty$, the Carter observer coincides with an inertial observer at rest with respect to the black hole. 
The Carter coframe $(e^\alpha)$ is the dual to the Carter tetrad (i.e. the unique set of 1-forms $(e^\alpha)$ such that $(e^\alpha)_\mu (e_\beta)^\mu = \delta^\alpha_{\ \, \beta}$). It is 
expressed in terms of the coframe $(\rd x^\alpha)$ associated to the Boyer-Lindquist coordinates by 
\begin{subequations}
\label{Carter_coframe}
\begin{align}
e^0 & = \sqrt{\frac{\Delta}{\Sigma}} \left(  \rd t - a \sin^2{\theta} \, \rd \varphi \right) \\
e^1 & = \sqrt{\frac{\Sigma}{\Delta}} \, \rd r \\
e^2 & = \sqrt{\Sigma} \, \rd \theta \\
e^3 & = \frac{\sin{\theta}}{\sqrt{\Sigma}} \left( - a  \, \rd t + (r^2 +a^2) \, \rd \varphi \right) .
\end{align}
\end{subequations}
We can now discuss the different types of interesting solutions belonging to the family (\ref{SOL}). 

\subsubsection{Asymptotically vanishing case: the electric and magnetic monopoles}

Let us first discuss the set of solution describing configurations for which $F\to 0$ for $r\to +\infty$. 
The first branch corresponds to $\beta =0$ in Eq.~\eqref{SOL}. In the following, we write
\be \label{alpha_cQ}
\alpha = -\frac{\cQ}{2M} ,
\ee
such that $\cQ$ encodes the strength of the field (actually the total electric charge, as we shall see below). Equation~\eqref{SOL} reduces then 
to $A^{(\alpha,0)}_{\mu} = \alpha \xi_{\mu}$, which yields
\begin{align} \label{A_alpha_0}
A^{(\alpha,0)} =  \frac{\cQ}{2M} \left(  1 - \frac{2 Mr}{\Sigma} \right) \rd t + \frac{a \cQ r \sin^2{\theta}}{\Sigma} \rd \varphi .
\end{align}
Note that the term $\frac{\cQ}{2M} \rd t$ is an exact differential and thus does not contribute to the electromagnetic field $F^{(\alpha,0)}$, so that a 4-potential equivalent 
to \eqref{A_alpha_0} is 
\begin{align} 
{A'}^{(\alpha,0)} = - \frac{\cQ r}{\Sigma} \rd t + \frac{a \cQ r \sin^2{\theta}}{\Sigma} \rd \varphi .
\end{align}
The associated Faraday tensor $F^{(\alpha,0)} = \rd A^{(\alpha,0)} = \rd {A'}^{(\alpha,0)}$ is given in
terms of the Carter coframe \eqref{Carter_coframe} by 
\begin{align}
\label{W1}
\boxed{F^{(\alpha,0)}  =  - \frac{\cQ (r^2 - a^2 \cos^2{\theta})}{\Sigma^2} \, e^0 \wedge e^1 + \frac{2 a \cQ r \cos{\theta}}{\Sigma^2} \, e^2 \wedge e^3}
\end{align}
A direct computation shows that this Maxwell solution has the electric charge $\cQ$, but no magnetic charge: 
\be
Q_{\rm e} = \frac{1}{4\pi} \oint_{S} \ast F^{(\alpha,0)}  = \cQ  \quad\mbox{and}\quad  Q_{\rm m} = \frac{1}{4\pi} \oint_S F^{(\alpha,0)}  = 0 .
\ee
The electric and magnetic field measured by the Carter observer are respectively
$E^{(\alpha,0)\mu} = F^{(\alpha,0)\mu}_{\phantom{(\alpha,0)\mu}\,  \nu} (e_0)^\nu$ and $B^{(\alpha,0)\mu} = - \ast F^{(\alpha,0)\mu}_{\phantom{(\alpha,0)\mu}\,  \nu} (e_0)^\nu$. 
Given that $(e_\alpha)$ is an orthonormal tetrad and $(e^\alpha)$ is its dual basis, we can read them directly on \eqref{W1}:
\begin{align}
E^{(\alpha,0)}  & = \frac{\cQ (r^2 - a^2 \cos^2{\theta})}{\Sigma^2} \, e_1 \\
B^{(\alpha,0)} & = \frac{2 a \cQ r \cos{\theta}}{\Sigma^2} \, e_1  .
\end{align}
They are both purely radial and vanish asymptotically.  When taking the non-rotating (Schwarzschild) limit, i.e. $a=0$, one finds  that 
\begin{align}
A^{(\alpha,0)} = \frac{\cQ}{2M} \left( 1 -\frac{2M}{r}\right) \rd t,  \qquad F^{(\alpha,0)} = -\frac{\cQ}{r^2} \,  e^0 \wedge e^1,
\qquad E^{(\alpha,0)} =  \frac{\cQ}{r^2} \, e_1, \qquad B^{(\alpha,0)} = 0 ,
\end{align}
which corresponds to an electric monopole. While this is the simplest solution one can construct, its electric charge makes it a less appealing model for the magnetosphere. Indeed, the plasma around an electrically charged magnetosphere would quickly discharge any charge excess so that the electrically neutral configuration is the natural assumption when considering the black hole magnetosphere on a long timescale. In order to consider a magnetosphere model without electric charge, one can either (i) combine it to the solution $F^{(0,\beta)}$ generated by the 
azimuthal Killing vector and appropriately choose the combination to make the total electric charge vanish, this is the well known Wald construction \cite{Wald:1974np}, or (ii) one can consider its dual solution 
$\ast F^{(\alpha,0)}$,
which instead has zero electric charge but a non-vanishing magnetic charge. 

Let us focus on option (ii), i.e. use the electromagnetic duality of the source-free Maxwell equations and consider the solution given by the Hodge dual of $F^{(\alpha,0)}$:  
\begin{align}
\label{KYsol}
\boxed{\tilde{F}^{(\alpha,0)}  = \ast F^{(\alpha,0)}  =   \frac{ 2 a \cP r \cos{\theta}  }{\Sigma^2}   e^0 \wedge e^1 + \frac{ \cP (r^2 - a^2 \cos^2{\theta})}{\Sigma^2} e^2 \wedge e^3} ,
\end{align}
where we have denoted $\cQ$ by $\cP$, to stress that this quantity will now represent a magnetic charge. 
The associated gauge-potential is given by (cf. Notebook~2 in App.~\ref{sage_notebooks})
\be \label{tilde_A_alpha_0}
\tilde{A}^{(\alpha,0)} = \frac{ \cP \cos{\theta}}{\Sigma} \left( a \rd t - (r^2+a^2) \rd \varphi\right) .
\ee
Anticipating on the next section, we notice that it satisfies the following interesting property:
\be
(r^2+a^2) \tilde{A}^{(\alpha,0)}_t + a \tilde{A}^{(\alpha,0)}_{\varphi} =0 .
\ee
This will reveal useful when studying the general conditions on the electrogeodesic motion. 
As expected, this dual configuration carries a magnetic charge and no electric charge, i.e. one has
\be
Q_{\rm e} = \frac{1}{4\pi} \oint_{S} * \tilde{F}^{(\alpha,0)}  = 0 \quad\mbox{and}\quad Q_{\rm m} = \frac{1}{4\pi} \oint_S \tilde{F}^{(\alpha,0)}  = \cP .
\ee
The electric and magnetic fields measured by the Carter observer are respectively
\begin{align}
\tilde{E}^{(\alpha,0)}  & = -  \frac{2 a \cP r \cos{\theta}}{\Sigma^2} \, e_1 \\
\tilde{B}^{(\alpha,0)} & =  \frac{\cP (r^2 - a^2 \cos^2{\theta})}{\Sigma^2} \, e_1 . 
\end{align}
They are both purely radial and vanishing for $r\to +\infty$. The solution \eqref{KYsol} describes thus a magnetic monopole of magnetic charge $\cP$
in the Kerr geometry. As we shall see, this solution admits an interesting interpretation in term of the Killing-Yano tensor of the Kerr geometry. In particular, it can be alternatively derived using the Penrose current construction that we discuss in the next section. 

Before presenting this construction, let us point that an interesting property of this magnetic monopole solution is that in the non-rotating (Schwarzschild) case, it reduces to 
\begin{align}
\tilde{F}^{(\alpha,0)} = \cP \sin{\theta} \, \rd \theta \wedge \rd \varphi ,
\end{align}
which is the well known non-rotating Michel solution extensively studied in the literature \cite{Michel}. This solution describes the simplest magnetic monopole in flat spacetime and on the Schwarzschild black hole. 
However, in the slow rotating case treated at linear order in $a$, one finds 
\be
\tilde{F}^{(\alpha,0)} \sim \cP \sin{\theta} \rd \theta \wedge \left( \rd \varphi - \frac{a}{r^2} \rd t\right) + \frac{2 a \cP \cos{\theta}}{r^3} \rd t \wedge \rd r + \cO(a^2) ,
\ee
which does not correspond to the slowly rotating Michel solution discussed in \cite{Michel}, i.e. $F^{M} = \sin{\theta} \rd \theta \wedge (\rd \varphi - \Omega \rd u)$ where $\rd u = \rd t + \rd r$ and $\Omega$ is a constant. Therefore, the solution (\ref{KYsol}) stands as the non-linear spinning generalization of the non-rotating Michel monopole solution. As we shall see, it enjoys special properties w.r.t to the separability of the electrogeodesic equation. 

Let us now stress that while the solution (\ref{tilde_A_alpha_0}) cannot describe an astrophysical situation because it carries a magnetic monopole, one can build out of this solution a more admissible split-monopole solution which is free from the magnetic charge. To do so, one can cut the solution at the equator such that 
\be
F_{\text{split}} = \left\{
    \begin{array}{ll}
        F_{\text{mono}} & 0 \leqslant \theta  < \pi/2 \\
        F_{\text{mono}} & \pi/2 \leqslant \theta  < \pi
    \end{array}
\right.
\ee
which automatically removes the magnetic charge when integrating over the sphere. In order to build such solution, one can consider the following simple modification of the gauge potential (\ref{tilde_A_alpha_0}) which reads
\begin{align}
\label{spli}
A_{\text{split}} = \frac{\cP a |\cos{\theta}|}{\Sigma} \rd t + \cP \left[ 1- \frac{(r^2+a^2) |\cos{\theta}|}{\Sigma} \right] \rd \varphi
\end{align}
The constant shift $\cP \rd \varphi$ allows us to avoid singularities on the axis of rotation. At large radius, $r \gg 1$, this solution behaves as $A = \cP (1 - |\cos{\theta}|) \rd \varphi + \cO(r^{-1})$ which reduces to the standard potential for the split monopole. The only price to pay is that the components of the Faraday tensor tangential to the equator can exhibit a jump which implies the presence of distributional current at the equator. However, as far as one focuses on the physics away from the equator, the Faraday tensor behaves the same except for a global change of sign from one hemisphere to the other. As we shall see, all the results we will present in the following remain valid for this split monopole solution (\ref{spli}) provided one shifts the definition of the angular momentum of the charged particles. 

\subsubsection{Asymptotically uniform case: the Wald solution}

Let us consider the second branch of solutions \eqref{SOL}, which leads to uniform magnetic fields following Wald's construction \cite{Wald:1974np}. 
The second branch corresponds to $\alpha=0$ in Eq.~\eqref{SOL} and we introduce the constant parameter $\cB$ such that 
\be \label{beta_cB}
\beta = \frac{\cB}{2}.
\ee
Since $A^{(0,\beta)}_{\mu} = \beta \chi_{\mu}$, there comes
\begin{align}
A^{(0,\beta)} = - \frac{ a \cB M r \sin^2{\theta}}{\Sigma} \, \rd t + \frac{\cB \left( r^4 + a^2 \Sigma + r^2 a^2 \cos^2{\theta}+ 2 a^2 M r \sin^2{\theta}\right) }{2\Sigma} \sin^2{\theta} \, \rd \varphi ,
\end{align}
which yields the following Faraday tensor (cf. Notebook~1 in App.~\ref{sage_notebooks}):
\begin{align}
F^{(0,\beta)} = & - \frac{a\cB\left[ M(r^2 - a^2 \cos^2\theta) + r \Sigma \right] \sin^2\theta}{\Sigma^2} \, e^0 \wedge e^1
  - \frac{a \cB \sqrt{\Delta} \sin\theta}{\Sigma} \, e^0 \wedge e^2
  + \frac{\cB \sqrt{\Delta} r \sin\theta}{\Sigma} \, e^1 \wedge e^3 \nonumber \\ 
  & + \frac{\cB \left[ r^4 + a^2(r^2 + 2 M r + \Delta \cos^2\theta) \right]\cos\theta}{\Sigma^2} \, e^2 \wedge e^3 . \label{W2}
\end{align}
Computing the electric and magnetic charges, one finds that this solution carries again only an electric charge, i.e.
\be
Q_{\rm e} =  \frac{1}{4\pi} \oint_{S} \ast F^{(0,\beta)}  = 2 a M \cB  \quad\mbox{and}\quad Q_{\rm m} =  \frac{1}{4\pi} \oint_S F^{(0,\beta)} = 0 . 
\ee
The electric and magnetic fields measured by the Carter observer are respectively 
\begin{align}
E^{(0,\beta)} & = \frac{a\cB\left[ M(r^2 - a^2 \cos^2\theta) + r \Sigma \right] \sin^2\theta}{\Sigma^2} \,  e_1 + \frac{a \cB \sqrt{\Delta} \sin\theta}{\Sigma} \, e_2  \\
B^{(0,\beta)} & = \frac{\cB \left[ r^4 + a^2(r^2 + 2 M r + \Delta \cos^2\theta) \right]\cos\theta}{\Sigma^2} \,  e_1 - \frac{\cB \sqrt{\Delta} r \sin\theta}{\Sigma} \, e_2 . 
\end{align}
The first remark is that, at large $r$, while the electric field decays to zero, the magnetic field asymptotes to a value which is independent of $r$, i.e one has
\be
B^{(0,\beta)} \sim \cB \left(  \cos{\theta} \, e_1 -  \sin{\theta} \, e_2 \right) .
\ee
This provides us with an asymptotically uniform magnetic field of amplitude $\cB$ parallel to the rotation axis. 

Following the Wald construction \cite{Wald:1974np}, it is then natural to consider the linear combination of the two solutions that allows for an \textit{electrically neutral} solution. Considering the linear combination 
\be
\label{W}
A^{\text{W}}_{\mu}  = \alpha  \xi_{\mu}  + \beta \chi_{\mu} ,
\ee 
one finds that the total electric charge is 
\be
Q^{\text{W}}_e =  \cQ + 2 a M \cB , 
\ee
where $\cQ$ and $\cB$ are related to $\alpha$ and $\beta$ by Eqs.~\eqref{alpha_cQ} and \eqref{beta_cB}. 
Hence, by choosing $\cQ= - 2 a M \cB$, one obtains an electrically neutral configuration: $Q^{\text{W}}_e =0$. 
This is by far the most studied solution to the Maxwell equations around a Kerr black hole. As we shall see, the electrogeodesic motion is not separable in this case. 

\subsubsection{Asymptotically growing case: a new toroidal solution}
\label{A3}

To conclude this section, we present a new way to construct an exact non-vacuum Maxwell solution on the Kerr geometry using the Killing vectors. Consider the following 
2-form $F$ that is the Hodge dual of the exterior product of the two Killing 1-forms:
\be
F_{\mu\nu} =  \;  \epsilon_{\mu\nu\rho\sigma} \xi^{\rho} \chi^{\sigma} .
\ee
Explicitly, it reads
\be
\label{newsol}
F =  \;  \Sigma \; \rd r \wedge \rd \theta \qquad \text{or equivalently} \qquad \boxed{F =  \Delta \sin{\theta} \; e^1 \wedge e^2} . 
\ee
It is direct to show that $F$ satisfies the non-vacuum Maxwell equations with an electric current given by (cf. Notebook~1 in App.~\ref{sage_notebooks})
\be
\label{newcurrent}
J =  \frac{2}{\Sigma} \left[ - \Delta \cos{\theta} \, \partial_r + (r-M) \sin{\theta} \, \partial_{\theta}\right] .
\ee
In the large $r$ limit, it reduces to $J \sim  2 \left( -  \cos{\theta} e_1 +  \sin{\theta} e_2 \right)$, i.e. $J$ becomes 
uniform and parallel to the rotation axis. 
With respect to the Carter observer, the electric and magnetic fields read
\begin{align}
E = 0 \quad\mbox{and}\quad B = \Delta \sin{\theta} \, e_3 .
\end{align}
Hence the solution \eqref{newsol} corresponds to a purely toroidal magnetic field in the Carter frame. 
Notice that the field is not uniform but growing at large $r$, so that if it is useful for a magnetosphere model, it can only be considered in a bounded region around the hole. Just as for the Wald solution, the electrogeodesic motion is not separable in this magnetosphere. 

Before discussing this point, let us first present the alternative construction of source-free Maxwell solutions based on the so called Penrose current.

\subsection{Maxwell solutions from Killing-Yano tensors}

In the previous section, we have reviewed two approaches to construct vacuum Maxwell solution on the Kerr geometry using the Killing vectors. As it turns out, one can also use the Killing-Yano tensors of the Kerr geometry to build vacuum and non-vacuum Maxwell solutions. As we shall see, some of these solutions can be related to each other using the Killing tower of the Kerr geometry. Therefore, we start by reviewing this key structure ruling the symmetries of the Kerr black hole. 

\subsubsection{The Killing tower}

\label{KTower}


Consider a metric which admits a rank-2 tensor $H$ satisfying
\begin{align}
\label{principal}
\nabla_{\mu} H_{\nu\alpha} = 2 g_{\mu[\nu} \xi_{\alpha]} , \qquad \nabla_{[\mu} H_{\nu\alpha]} =0, \qquad \xi_{\alpha} = \frac{1}{3} \nabla_{\mu} H^{\mu}{}_{\alpha} .
\end{align}
The tensor $H$ is a closed non-degenerate conformal Killing-Yano (CCKY) 2-form. Following \cite{Krtous:2006qy, Kubiznak:2006kt, Krtous:2008tb}, we call this object the principal tensor.
Now, taking the dual of the principal tensor, one can build a rank-$2$ antisymmetric tensor satisfing
\be
Y_{\mu\nu} = \frac{1}{2} \epsilon_{\mu\nu\rho\sigma} H^{\rho\sigma},  \qquad \nabla_{(\mu} Y_{\nu)\alpha} =0 ,
\ee
which corresponds to a rank-$2$ Killing-Yano (KY) tensor. 
From this KY tensor, one can construct a rank-$2$ (symmetric) tensor satisfying 
\be
\label{KT}
K^{\mu\nu} = Y^{\mu\alpha} Y_{\alpha}{}^{\nu}, \qquad \nabla_{(\alpha} K_{\mu\nu)} =0 ,
\ee
which thus provides the Killing tensor for the underlying geometry.
Finally, using the principal tensor and the Killing tensor, one can further construct two vectors $\xi$ and $\chi$ such that 
\begin{align}
\label{KV1}
\xi^{\mu}  & = \frac{1}{3} \nabla_{\alpha} H^{\alpha \mu}  , \qquad  \text{such that } \qquad \nabla_{(\mu} \xi_{\nu)} = 0 \\
\label{KV2}
\chi^{\mu}  & = - K^{\mu}{}_{\alpha} \xi^{\alpha}  , \qquad  \text{such that } \qquad \nabla_{(\mu} \chi_{\nu)} = 0 .
\end{align}
This provides two Killing vectors for the underlying metric. In summary, any metric equipped with a principal tensor is automatically equipped with a rank-$2$ KY tensor and its associated Killing tensor while it also enjoys two isometries. This structure is called the Killing tower, as one can derive all the Killing objects of the metric starting from one single object, the principal tensor. It was identified and described first in the context of higher dimensional black holes in \cite{Krtous:2006qy, Kubiznak:2006kt, Krtous:2008tb}. See \cite{Frolov:2017kze} for a review. 

As it turns out, the Kerr geometry enjoys such a structure. Concretely, the principal tensor and the KY tensor of the Kerr geometry are given by (cf. Notebook~1 in App.~\ref{sage_notebooks})
\begin{align}
\label{Hp}
H = r e^0 \wedge e^1 + a \cos{\theta} \; e^2 \wedge e^3 \\ 
Y = a \cos{\theta} \; e^0 \wedge e^1 - r e^2 \wedge e^3 
\end{align}
One can easily check that, using the definitions \eqref{KV1}--\eqref{KV2}, one recovers the two commuting Killing vectors \eqref{Kerr_Killing_vectors}. 
Having reviewed these hidden Killing objects of the Kerr geometry, we can now discuss two alternatives ways to construct exact solutions of the Maxwell equations on the Kerr geometry.

\subsubsection{The Killing-Maxwell system from the principal tensor}

Consider the principal tensor $H$ of the Kerr geometry given by (\ref{Hp}). It provides an exact solution of the non-vacuum Maxwell equation known as the Killing-Maxwell (KM) solution  \cite{Carter:1987id}. 
To see this, notice first that by definition, it is a closed form: $\rd H =0$. Let us introduce the associated KM Faraday tensor by 
\be
F^{\text{KM}} = \frac{1}{3} H  = \frac{1}{3} \left(  r e^0 \wedge e^1 + a \cos{\theta} \; e^2 \wedge e^3\right) .
\ee
It can be derived from the 4-potential
\be
A^{\text{KM}} =  \frac{1}{6} \left[- \left( r^2 - a^2 \cos^2{\theta}\right)\rd t + (r^2 \sin^2{\theta} + a^2 \cos^2{\theta}) a \rd \varphi \right]
\ee
Then, using the last equation of (\ref{principal}), we get
\be
\nabla_{\nu} F_{\text{KM}}^{\mu\nu} = 4\pi J^{\mu} \qquad \text{with} \qquad J = \frac{1}{4\pi} \partial_t .
\ee
Hence $F^{\text{KM}}$ fulfills the  Maxwell equations with an electric 4-current given by the asymptotically timelike Killing vector $\xi  = \partial_{t}$, up to a $4\pi$ factor. 
It is remarkable that the fundamental Killing object of the Killing tower provides without any further condition an exact non-vacuum solution of the test Maxwell equation. This system was studied by Carter \cite{Carter:1987id}, who showed that even more remarkably, it provides one example of a magnetosphere for the Kerr black hole which preserves the existence of the Carter constant for the electrogeodesic motion. The same construction was later studied by Krtous in \cite{Krtous:2007xg} to build Maxwell solutions on higher dimensional rotating black holes. 

\subsubsection{The Kerr magnetic monopole from the Penrose current}

Another approach to construct vacuum Maxwell solutions in a metric possessing a KY tensor $Y$ was introduced by Penrose in \cite{Penrose:1982wp}. Consider the following antisymmetric rank-$2$ tensor
\be
\label{Penrose}
F_{\mu\nu} = \alpha \; R_{\mu\nu\alpha\beta} Y^{\alpha\beta} .
\ee
Taking its divergence, one has
\begin{align}
\nabla_{\mu} F^{\mu\nu} & = R^{\rho\sigma\mu\nu}\nabla_{\mu} Y_{\rho\sigma} +  Y^{\rho\sigma} \nabla_{\mu} R_{\rho\sigma}{}^{\mu\nu} \nonumber \\
& = -  R^{\nu\mu\rho\sigma} \nabla_{[\mu} Y_{\rho\sigma]}   + 
Y^{\rho\sigma} \cancel{\nabla_{\mu} R_{\rho\sigma}{}^{\mu\nu} } \nonumber \\
& = -  R^{\nu[\mu\rho\sigma]}  \nabla_{\mu} Y_{\rho\sigma}  \nonumber \\
& = 0 .
\end{align}
Here, we have used the cyclic properties of the Riemann tensor, i.e. $R_{\mu[\nu\rho\sigma]} =0$, the identity $\nabla_{\mu} R_{\rho\sigma}{}^{\mu\nu} = 0$, which holds only in vacuum (i.e. when $R_{\alpha\beta} = 0$), and the definition of the KY tensor given by
\be
\nabla_{(\mu} Y_{\nu)\alpha} =0 \qquad \nabla_{\mu} Y_{\nu\alpha} = \nabla_{[\mu} Y_{\nu\alpha]} =  0 \qquad \nabla_{\mu} Y^{\mu}{}_{\alpha} =0 .
\ee
Taking the Hodge dual, one finally obtains 
\be
\nabla_{\mu} F^{\mu\nu} = 0 \quad\mbox{and} \quad \nabla_{\mu} *F^{\mu\nu} = 0 ,
\ee
where $*F_{\mu\nu} := \tfrac{1}{2} \epsilon_{\mu\nu\rho\sigma} F^{\rho\sigma}$ is the dual tensor. Interestingly, applying this to the Kerr geometry and using $\alpha = -\cP/(2M)$
in \eqref{Penrose}, 
one obtains\footnote{See Notebook~1 in App.~\ref{sage_notebooks} for the computation.}
\begin{align}
\label{KYsol2}
F & =  \frac{ 2 a \cP r \cos{\theta}  }{\Sigma^2}   e^0 \wedge e^1 + \frac{\cP (r^2 - a^2 \cos^2{\theta})}{\Sigma^2} e^2 \wedge e^3 ,
\end{align}
which reproduces the solution (\ref{KYsol}) discussed in the previous section. Therefore, the Penrose current of the Kerr spacetime coincides with the dual solution of the Wald construction based solely on the stationary Killing vector $\xi$. This concludes the review on the construction of exact Maxwell solutions from the underlying symmetries. Let us now discuss how to study the motion of charges particles in such magnetospheres. 


\section{Motion of charged particles in the Kerr magnetosphere}

\label{B}

In this section, we review several key properties for the electrogeodesic motion. After presenting the phase space, we discuss the general conditions that the energy and angular momentum of an electrogeodesic have to satisfy in the Kerr geometry. Then, we review the key geometrical property ensuring that the Carter constant is preserved, and thus that the electrogeodesic equations are separable. Finally, we rederive the first order electrogeodesic equations. 


In what follows, we shall consider a Maxwell field whose gauge-potential is restricted to the form
\be
\label{A}
A_{\mu} \rd x^{\mu} = A_{t}(r,\theta) \, \rd t + A_{\varphi} (r,\theta)\,  \rd \varphi .
\ee
w.r.t the Boyer-Lindquist coordinates.
This covers in particular the case of the test Kerr magnetic monopole which will be the central magnetosphere model we will study.

\subsection{Lagrangian and phase space}

\label{B1}

Consider a particle with mass $m$ and electric charge $q$ moving on a curved geometry described by the metric $g$ filled with a gauge-potential $A$. From the beginning, we exclude the case 
$m =0$ since there are no known charged particles with zero mass.
The dynamics of this charged particle is encoded in the Lagrangian 
\begin{align}
\cL = \frac{1}{2} m g_{\mu\nu} u^{\mu} u^{\nu} + q A_{\mu} u^{\mu} \qquad \text{with} \qquad u^{\mu} = \frac{\rd x^{\mu}}{\rd \lambda} = \dot{x}^{\mu} ,
\end{align}
where $\lambda$ is the proper time along the electrogeodesic and we have denoted $x^{\mu}(\lambda)$ the position of the particle and $u^{\mu} = \rd x^{\mu}/\rd \lambda$ its $4$-velocity. 
The equations of motion reads
\begin{align}
\label{elec}
m u^{\alpha} \nabla_{\alpha} u^{\mu} = q F^{\mu}{}_{\nu} u^{\nu} . 
\end{align}
The particle is no longer in free fall but moves under the action of both the geometry and the Lorentz force $F^{\mu}{}_{\nu} u^{\nu}$.

Switching to the Hamiltonian formulation, the canonical momentum derived from the Lagrangian is given by
\be
\label{genmom}
P_{\mu} = \frac{\delta \cL}{\delta \dot{x}^{\mu} } = m g_{\mu\nu} \dot{x}^{\nu} + q A_{\mu} .
\ee
It follows that the canonical brackets are given 
\be
\{ x^{\mu}, x^{\nu}\} = 0 \qquad \{ x^{\mu} , P_{\nu}\} = \delta^{\mu}{}_{\nu} \qquad \{ P_{\mu} , P_{\nu}\} = 0
\ee
The Hamiltonian $\cH$ takes the form 
\be
\label{massrel}
\cH = \frac{1}{2}g^{\mu\nu} \left( P_{\mu} -  q A_{\mu} \right) \left( P_{\nu} - q A_{\nu} \right) = -\frac{m^2}{2}
\ee
From this expression, it is natural to introduce another set of canonical variables and in particular the dressed momentum
\be \label{p_P_qA}
p_{\mu} = P_{\mu} - q A_{\mu}
\ee
which is related to the $4$-velocity of the electrogeodesic flow through $p_{\mu} = m u_{\mu}$.  This dressed momentum satisfies the following brackets
\be
 \{ x^{\mu} , p_{\nu}\} = \delta^{\mu}{}_{\nu}  \qquad \{ p_{\mu} , p_{\nu}\} = q F_{\mu\nu}
\ee
While this momentum is obviously not canonical, it will reveal useful when discussing the conditions to preserve the Carter constant. In term of this momenta, the Hamiltonian takes the usual form 
\be
H = \frac{1}{2}g^{\mu\nu} p_{\mu} p_{\nu}
\ee
such that one has $\{ p_{\mu}, H\} = 0$, i.e. $p_{\mu}$ is indeed conserved along the flow. 
Therefore, the presence of the Maxwell field introduces a distinction between the canonical momentum $P_{\mu}$, which serves as computing the conserved charges along the flow in GR, and the momentum $p_{\mu}$ associated to the $4$-velocity of the flow $ u_{\mu}$.  As we shall see, this will have interesting consequences for the trajectories of charged particles.
We can now discuss the properties of the electrogeodesics.

\subsection{General constraints on charged particles}

\label{B2}

A set of conditions can be derived for the geodesic motion of neutral particles on the Kerr geometry. They descend from the basic requirement that the $4$-velocity $u^{\mu}$ is a timelike vector. They can be expressed in terms of the energy and the angular momentum of the neutral massive particle. As they have important consequences for the geodesic motion, our first goal here is to discuss how these constraints are modified for electrogeodesics. 

To start with, let us first define the conserved energy $E$ and conserved angular momentum $L$ of the charged particle. As usual, the Kerr geometry being stationary and axisymmetric, one can use its time translation Killing vector $\xi = \partial_t$ and its azimuthal Killing vector $\chi = \partial_{\varphi}$ to build 
$E$ and $L$. Using the canonical momentum (\ref{genmom}), one obtains
\begin{align}
\label{E}
 - E & =  \xi^{\mu} P_{\mu} = P_t =  g_{tt} p^t + g_{t\varphi} p^{\varphi} + q A_t \\
 \label{L}
  L & = \chi^{\mu} P_{\mu} =  P_{\varphi} =  g_{\varphi\varphi} p^{\varphi}  + g_{\varphi t} p^t  + q A_{\varphi}
\end{align}
which can be recast into
\begin{align}
\label{E}
E  & = \left( 1 - \frac{2M r}{\Sigma}\right) p^t + \frac{2aM r \sin^2{\theta}}{\Sigma} p^{\varphi}  - q A_t\\
\label{L}
L & =  -  \frac{2aM r \sin^2{\theta}}{\Sigma}  p^t +\left( r^2 +a^2 + \frac{2 a^2 M r \sin^2{\theta}}{\Sigma} \right) \sin^2{\theta} \; p^{\varphi}  + q A_{\varphi}
\end{align}
$E$ and $L$ provide the corrected notion of energy and momenta in the presence of the Maxwell field. 

Let us now review the different conditions on the electrogeodesics. 
\begin{itemize}
\item \textbf{First constraint:} Outside the ergoregion, i.e. for\footnote{$r_{\rm ergo}(\theta) = M + \sqrt{M^2 - a^2\cos^2\theta}$ defines the boundary of the ergoregion, i.e. the ergosphere.}  $ r_{\rm ergo}(\theta) < r < +\infty$, 
the Killing vector $\xi = \partial_t $ is by definition
future-directed timelike. Therefore, its scalar product with the (future-directed timelike) $4$-velocity $u^{\mu}$ of the electrogeodesic must be negative:
\be
g(\xi, u) < 0 ,
\ee
which translates into
\be
\frac{P_t - q A_t}{m} = - \frac{E + q A_t}{m} < 0 .
\ee
It follows that outside the ergoregion, the energy of the particle is constrained to satisfy
\be
\label{cond1}
\boxed{E > -q A_t } .
\ee
When $q=0$,  one recovers the well known result that neutral particles admit only positive energy state outside the ergosphere, i.e. $E > 0$ for $r > r_{\rm ergo}(\theta)$. 


\item \textbf{Second constraint :} The second constraint can be formulated in term of the scalar product of the vector $N^\mu = - \nabla^\mu t$ with the particle's $4$-velocity 
$u^\mu$. Since both vectors are future-directed timelike in the black hole exterior, one must have
\be
g(N, u)  <0 \qquad \rightarrow \qquad u^t > 0 ,
\ee
which can be translated into
\begin{align}
u^t & = g^{tt} u_t + g^{t\varphi} u_{\varphi} \\
& = \frac{1}{m \Sigma \Delta} \left[ \left[ (r^2+a^2) \Sigma + 2 a^2 M r \sin^2{\theta} \right](E + q A_t) -  2 a M r (L - q A_{\varphi})  \right]> 0 .
\end{align}
In the region $ r_{+} < r < \infty$, one has $\Delta > 0$. Therefore, in the outer region of communication of the Kerr geometry, this condition can be simplified to
\be
\label{cond2}
\boxed{(r^2 +a^2) \Sigma E - 2 a M r (L - a E \sin^2{\theta}) > 2 q a M r (a A_t \sin^2{\theta} - A_{\varphi})} .
\ee
\item \textbf{Third constraint :} A third set of conditions can be obtained from the $4$-velocity of the Carter frame, i.e. $e_0$. 
This vector being future-directed timelike its scalar product with $p^\mu$ must be negative: 
\be
g(e_0, p) <0 . 
\ee
Writing $g(e_0, p) = (e_0)^\mu p_\mu$ and using Eq.~\eqref{e_0_BL} for $(e_0)^\mu$ and Eq.~\eqref{p_P_qA} for $p_\mu$, with $P_t = -E$ and $P_\varphi = L$, 
one gets
\be
\label{cond3}
\boxed{(r^2 +a^2) E  - a L  >  -q  \left[ (r^2 +a^2) A_t + a  A_{\varphi}\right] } .
\ee
In the case of neutral particles where $q =0$, this condition implies two important consequences. First, for negative energy states with $E \leqslant 0$, which exist only in the ergosphere, 
the neutral particle must have $L <0$; hence it follows a retrograde orbit. Moreover, for neutral particles with $L =0$, one has automatically that $E > 0$. We see that (\ref{cond3}) does not implies these properties anymore, except if the magnetosphere is such that $(r^2 +a^2) A_t + a  A_{\varphi} =0$. As we shall, see this is the case of the magnetic monopole solution (\ref{KYsol}).
\end{itemize}
We can now turn to the conditions for separability of the electrogeodesic equations. 

\subsection{Conditions for preserving the Carter constant}

\label{B3}

An important property of the geodesic motion on the Kerr spacetime is that it exhibits a set of four constants of motion which are in involution. Two of them are the energy $E$, the angular momentum $L$ which descend from the time and azimuthal isometries of the Kerr geometry. The third one is the mass $m$ of the particle. However, the fourth constant of motion was found by Carter in \cite{Carter:1968rr} and descends from the existence of the hidden symmetry of the Kerr geometry giving rise to the Killing tower reviewed in Section~\ref{KTower}. Concretely, for a neutral test particle with momenta $p_{\mu}$, the Killing tensor (\ref{KT}) gives rise to a constant $\cK =K_{\mu\nu} p^{\mu} p^{\nu}$ along the geodesic flow, i.e. $D\cK =0$, known as the Carter constant, where $\cD = u^{\mu} \nabla_{\mu}$ is the derivative along the geodesic flow. 

It is therefore natural to wonder under which conditions the Carter constant is preserved for the electrogeodesic motion in a given magnetosphere of the Kerr geometry. This task was explored  in \cite{Carter:1987id} and clarified systematically in \cite{Visinescu:2009rm}. See also \cite{Igata:2010ny, Igata:2010bu, Obukhov:2020sjc, Obukhov:2021ggp, Obukhov:2023uxi}. Let us recall that the phase space of the electrogeodesic motion can be written as 
\be
 \{ x^{\mu} , p_{\nu}\} = \delta^{\mu}{}_{\nu}  \qquad \{ p_{\mu} , p_{\nu}\} = \kappa F_{\mu\nu} \qquad \text{with} \qquad p_{\mu} = P_{\mu} - q A_{\mu}
\ee
with the Hamiltonian of the form
\be
H = \frac{1}{2}g^{\mu\nu} p_{\mu} p_{\nu} .
\ee 
Consider now the following generalized Carter constant defined by
\be
\cK = K^{\mu\nu} p_{\mu} p_{\nu} .
\ee
Then, the Hamiltonian evolution of the function $\cK$ is given by
\begin{align}
\{ \cK , H \} & = \left( \partial_{\rho} K^{\mu\nu}\right) g^{\alpha\beta} p_{\mu} p_{\nu} p_{\beta} \{ x^{\rho}, p_{\alpha}\} -  K^{\mu\nu} \left( \partial_{\rho} g^{\alpha\beta} \right) p_{\nu} p_{\alpha} p_{\beta} \{ x^{\rho}, p_{\mu}\} + K^{\mu\nu} g^{\alpha\beta} p_{\nu} p_{\beta} \{ p_{\mu}, p_{\alpha}\} \nn \\
& = \frac{1}{6}\nabla^{(\alpha} K^{\mu\nu)} p_{\alpha} p_{\mu} p_{\nu} + F_{\mu}{}^{(\alpha} K^{\nu)\mu} p_{\alpha}  p_{\nu} ,
 \end{align}
 which splits into a term cubic in the momenta and a second term quadratic in the momenta. 
 By definition of the Killing tensor (\ref{KT}), the first term vanishes reducing the bracket to
 \begin{align}
\{ \cK , H \} & =  F_{\mu}{}^{(\alpha} K^{\nu)\mu} p_{\alpha}  p_{\nu} ,
 \end{align}
which can be interpreted as the anomaly induced by the presence of the magnetosphere. Therefore, the condition for the Carter constant to remain a constant of motion of the electrogeodesic motion is given by
\be
\{ \cK , H \}  = 0  \qquad \text{if} \qquad F_{\mu}{}^{(\alpha} K^{\nu)\mu}  =0 .
\ee
Since the Killing tensor descends from the KY $2$-form of the Kerr geometry, this condition can be written as follows
\begin{align}
F_{\mu}{}^{(\alpha} K^{\nu)\mu}  & = ( F_{\mu}{}^{\alpha} Y^{\beta \mu} ) Y^{\nu}{}_{\beta}  + (F_{\mu}{}^{\nu} Y^{\beta\mu}) Y^{\alpha}_{\beta} .
\end{align} 
Let us now impose that
\be
\label{COND}
S^{\alpha\beta} = F_{\mu}{}^{[\alpha} Y^{\beta] \mu} = 0 .
\ee
Plugging this relation in the previous expression, one finds 
\begin{align}
F_{\mu}{}^{(\alpha} K^{\nu)\mu}  & = ( F_{\mu}{}^{\beta} Y^{\alpha \mu} ) Y^{\nu}{}_{\beta}  + (F_{\mu}{}^{\beta} Y^{\nu\mu}) Y^{\alpha}_{\beta}  = Y^{\alpha\beta} Y^{\nu\mu} \left( F_{\beta\mu} + F_{\mu\beta} \right) = 0 .
\end{align} 
Therefore, the condition (\ref{COND}) is sufficient to ensure that $F_{\mu}{}^{(\alpha} K^{\nu)\mu} = 0$, as first noticed in \cite{Visinescu:2009rm}.  This condition selects the models of Kerr magnetosphere that preserve the Carter constant.  Notice that while we demand here that the Carter constant be preserved, it is in principle possible that for some magnetosphere models, the electrogeodesic is separable for a corrected version of the Carter constant. The condition (\ref{COND}) only applies to the former case. 

We can now check which solution for the Kerr magnetosphere satisfies this peculiar condition. Let us examine each magnetosphere model discussed in the previous section. 
\begin{itemize}
\item \textbf{Killing-Maxwell solution}: It is direct to check that
\be
F_{\alpha\beta} = H_{\alpha\beta}, \qquad S^{\alpha\beta} = H_{\mu}{}^{[\alpha} Y^{\beta] \mu}  = 0 .
\ee  
To our knowledge, together with the Kerr-Newman black hole magnetosphere, this is the only Maxwell solution satisfying that property.
This was shown by Carter in \cite{Carter:1987id}. 
\item \textbf{The Kerr magnetic monopole} (and Penrose current): Consider now the magnetosphere built from the Penrose current (or the dual timelike Killing vector solution). One has
\be
F_{\mu\nu} = R_{\mu\nu\rho\sigma} Y^{\rho\sigma},  \qquad S^{\alpha\beta} = R_{\mu}{}^{[\alpha}{}_{\rho\sigma} Y^{\rho\sigma} Y^{\beta]\mu} = 0 .
\ee 
This is expected as this magnetosphere is the magnetic subsector of the Kerr-Newman magnetosphere.
This opens the door to analytically study the electrogeodesic motion in this Kerr magnetosphere which is considered test, i.e. without backreaction. This will be discussed in the next section.
\item \textbf{Wald solution}: Consider now the Wald solution.  One finds\footnote{See Notebook~1 in App.~\ref{sage_notebooks} for the computation.} that 
$S^{\alpha\beta} \neq 0$ in this case: the 2-form $S_{\alpha\beta}$ metric-dual to $S^{\alpha\beta}$ is
\be
    \underline{S} = \frac{\cB}{2} \sqrt{\Delta} \sin\theta \, e^1 \wedge e^2 . 
\ee
This explains why the electrogeodesic motion is not separable on the Wald solution.
\end{itemize}
Having reviewed the conditions for a magnetosphere to preserve the Carter constant, and checked that the Kerr monopole solution belongs to this very special class, we naturally turn towards writing the first order electrogeodesic equations.

\subsection{Electrogeodesic motion in the Kerr monopole magnetosphere: Separability}

\label{B4}

The main goal of this section is to review the different steps to separate the electrogeodesic equations within the Kerr test monopole magnetosphere. The first order electrogeodesic equations we shall derive are a subset of the electrogeodesic equations of the Kerr-Newman black hole \cite{Hackmann:2013pva} where the electric charge is set to zero and the magnetic charge is kept arbitrary.

\subsubsection{Separability of the radial and the polar motion}

In order to separate the radial equation, we first consider the Carter constant given by
\be
\cK = K_{\mu\nu} p^{\mu} p^{\nu} ,
\ee
where $K_{\mu\nu}$ is the Killing tensor of the Kerr geometry which can be decomposed as
\begin{align}
K_{\alpha\beta} = (r^2+a^2) (k_{\alpha} \ell_{\beta} + \ell_{\alpha} k_{\beta}) + r^2 g_{\alpha\beta} .
\end{align}
Here $k^{\alpha}$ and $\ell^{\alpha}$ are the null vectors corresponding to the principal null directions of the Kerr geometry; they are given explicitly by 
\begin{align}
k^{\alpha} \partial_{\alpha} & = \frac{r^2+a^2}{\Delta} \partial_t - \partial_r + \frac{a}{\Delta} \partial_{\varphi} \\
\ell^{\alpha} \partial_{\alpha} & = \frac{1}{2} \partial_t + \frac{\Delta}{2(r^2+a^2)} \partial_r + \frac{a}{2(r^2+a^2)} \partial_{\varphi}
\end{align}
A direct computation of the Carter constant gives 
 \begin{align}
 \label{ModCart}
 \cK & = \frac{1}{\Delta} \left[ \left( (r^2+a^2) (E + q A_t) - a (L- q A_{\varphi})\right)^2 - \Sigma^2 (p^r)^2\right] - r^2 m^2 \nonumber \\
 & = \frac{1}{\Delta} \left[ \left( (r^2+a^2) E - a L + { q((r^2+a^2) A_t + a  A_{\varphi})}\right)^2 - \Sigma^2 (p^r)^2\right] - r^2 m^2 ,
 \end{align}
 such that one obtains a non-trivial contribution from the magnetosphere encoded in the last term. 
 Interestingly, focusing on the Kerr monopole, one finds that this term vanishes:
 \be
 \label{COND22}
 (r^2+a^2) A_t + a A_{\varphi} =0 ,
 \ee
 so that the Carter constant takes the same form as for a geodesic:
  \begin{align}
 \label{ModCart}
 \cK  & = \frac{1}{\Delta} \left[ \left( (r^2+a^2) E - a L\right)^2 - \Sigma^2 (p^r)^2\right] - r^2 m^2 .
 \end{align}
 Since this constant depends only on the radial component $p^r$, it allows one to algebraically deduce the first order radial equation, which reads 
\begin{align}
\label{rad}
 \Sigma \; p^r = \Sigma m \frac{\rd r}{\rd \lambda}  =  \sqrt{R(r)}  \qquad \text{with} \qquad R(r) = \left[ (r^2+a^2) E - a L\right]^2 - \Delta (r^2 m^2 + \cK) .
\end{align}
This equation reproduces that of the geodesic case; notably, the l.h.s. only involves functions of the radial coordinate $r$.  Nevertheless, one has to keep in mind that now, it involves a corrected notion of the energy and angular momentum.

Now, we can obtain the first order equation for the polar motion using both the Carter constant and the mass relation $m^2 = - g_{\mu\nu} p^{\mu} p^{\nu}$. Writing this explicitly, we find
\begin{align}
\label{m2_prov}
m^2 = - g^{\mu\nu} \left(P_{\mu} - q A_{\mu} \right) \left( P_{\nu} - q A_{\mu} \right) =  - g^{\mu\nu} P_{\mu} P_{\nu} + 2 q g^{\mu\nu} P_{\mu} A_{\nu} - q^2 g^{\mu\nu} A_{\mu} A_{\nu} .
\end{align}
Since $A_r = A_{\theta} = 0$ so that $P_r =p_r$ and $P_{\theta} = p_{\theta}$, the first term will reproduce the standard term for the geodesic. The challenge lies in the last two terms which need to provide separable form. Concretely, they are given by
\begin{align}
g^{\mu\nu} P_{\mu} A_{\nu} & = - E \left( g^{tt} A_t + g^{t\varphi} A_{\varphi} \right) + L \left( g^{t\varphi} A_t + g^{\varphi\varphi} A_{\varphi} \right) 
 =  \frac{\cP \cos{\theta}}{\Sigma} \left( a E - \frac{L}{\sin^2{\theta}}\right)\\
g^{\mu\nu} A_{\mu} A_{\nu} & = \frac{\cP^2}{\Sigma} \frac{\cos^2{\theta}}{\sin^2{\theta}} .
\end{align}
Equation~\eqref{m2_prov} becomes then 
\begin{align}
m^2 & = -  \frac{1}{\Delta} \left( r^2 +a^2 + \frac{2 a^2 M r \sin^2{\theta}}{\Sigma} \right) E^2 - \frac{4 a M r }{\Sigma \Delta} E L - \frac{1}{\Delta} \left( 1 - \frac{2M r}{\Sigma}\right) \frac{L^2}{\sin^2{\theta}} \nn \\
& \qquad + \frac{2q \cP \cos{\theta}}{\Sigma} \left( a E - \frac{L}{\sin^2{\theta}}\right) 
   -  \frac{q^2 \cP^2}{\Sigma} \frac{\cos^2{\theta}}{\sin^2{\theta}} \nn \\
& \qquad - \frac{\Sigma}{\Delta} (p^r)^2 - \Sigma (p^{\theta})^2 \label{m2_pr_ptheta} .
\end{align}
Upon using (\ref{rad}) to rewrite $p^r$ as
  \begin{align}
 \label{ModCart}
\frac{\Sigma}{\Delta} (p^r)^2 = \frac{1}{\Delta} \left[ (r^2+a^2) E - a L\right]^2 - (r^2 m^2 + \cK)
 \end{align}
and injecting in Eq.~\eqref{m2_pr_ptheta}, one obtains after a little algebra that 
\begin{align}
\Sigma \; p^{\theta} =  \Sigma m \frac{\rd \theta}{\rd \lambda}  = \pm \sqrt{\Theta(\theta)} 
\end{align}
with the latitudinal potential $\Theta(\theta)$ defined by
\begin{align}
  \Theta(\theta) = \cK - m^2 a^2 \cos^2{\theta} -  \left( \frac{L +  q \cP \cos\theta}{\sin{\theta}} - a E \sin{\theta} \right)^2 .
\end{align}
Thus, contrary to the radial motion, the polar electrogeodesic motion is corrected w.r.t the geodesic motion. However, it also involves only functions of the polar angle $\theta$. In conclusion, the property (\ref{COND}) provides us with a Carter-like constant which, together with the mass, can be used to derive the first order equations for the radial and polar motion. Both equations are separable. Since the radial equation remains the same, this already implies that the different radial equilibrium positions for the charged particles will remain the same as the one of their neutral counterpart. On the contrary, the polar equilibrium positions will be modified. We shall come back on this point in the following. 

\subsubsection{Separability of the equation for the time and azimuthal motion}

Let us now turn to the equations for $\dot{t}$ and $\dot{\varphi}$. The expressions for the energy and the angular momentum can be written as 
\begin{align}
\mat{cc}{ - E - q A_t \\   L - q A_{\varphi}} & =  \mat{cccc}{ g_{tt} & g_{t\varphi}  \\  g_{\varphi t} &  g_{\varphi\varphi}} \mat{cc}{p^t \\ p^{\varphi}} .
\end{align}
Inverting this system, one obtains
\be
\mat{cc}{ p^t \\   p^{\varphi} }  = \frac{1}{g_{tt} g_{\varphi\varphi} - g^2_{t\varphi}}  \mat{cccc}{ g_{\varphi\varphi}  & - g_{t\varphi}  \\    -  g_{\varphi t} & g_{tt }}  \mat{cc}{ - E - q A_t \\   L - q A_{\varphi}}  ,
\ee
which provides us with the two following equations
\begin{align}
 m \frac{\rd t }{\rd \lambda}& =  - \frac{g_{t\varphi}  L + g_{\varphi\varphi} E}{ g_{tt} g_{\varphi \varphi} - g^2_{t\varphi} } - q \frac{g_{\varphi\varphi} A_{t} - g_{t\varphi} A_{\varphi}}{ g_{tt} g_{\varphi \varphi} - g^2_{t\varphi}} \\
  m \frac{\rd \varphi }{\rd \lambda} & 
= \frac{g_{tt}  L+ g_{t \varphi} E}{g_{tt} g_{\varphi \varphi} - g^2_{t\varphi} } -  q \frac{g_{tt} A_{\varphi} - g_{t \varphi} A_t}{ g_{tt} g_{\varphi \varphi} - g^2_{t\varphi}} .
\end{align} 
When the magnetosphere is turned off, i.e. when $q =0$, one recovers the standard equations for the time and azimuthal motions of the neutral particles on the Kerr geometry. When $q \neq 0$, the equations inherit new terms. Using that
\be
g_{tt} g_{\varphi \varphi} - g^2_{t\varphi} = - \Delta \sin^2{\theta}
\ee
one can show that these new terms read
\begin{align}
q \frac{g_{\varphi\varphi} A_{\varphi} - g_{t\varphi} A_t}{ - \Delta \sin^2{\theta}} & = - q  \frac{ a \cP \cos{\theta}}{\Sigma} \\
q \frac{g_{tt} A_{\varphi} - g_{t \varphi} A_t}{ - \Delta \sin^2{\theta}} & = - q \frac{ \cP \cos{\theta}}{\Sigma \sin^2{\theta}} .
\end{align}
Upon an overall factor of $1/\Sigma$, both contributions only involve a function of the polar angle. Upon using the radial and polar equations, this ensures that the equations for $\dot{t}$ and $\dot{\varphi}$ are indeed separable. They are explicitly given by 
\begin{align}
\Sigma m \frac{\rd t }{\rd \lambda} &  = \frac{1}{\Delta} \left[ (r^2+a^2)^2 E - 2 M a r L \right]  - a^2 E \sin^2{\theta}  + a q \cP  \cos{\theta} \\
 \Sigma m \frac{\rd \varphi }{\rd \lambda} & =  \frac{L}{\sin^2{\theta}} + \frac{a  }{\Delta} \left(  2 M r E - a L \right) + \frac{q \cP \cos{\theta}}{ \sin^2{\theta}} .
\end{align}
This concludes the review of the proof of the separability of the electrogeodesic equations in the test magnetic monopole on the Kerr geometry.

\subsection{Integral solution to the electrogeodesic equation}

We are now in position to present the exact analytic solution to the electrogeodesic equation in the Kerr magnetic monopole. For clarity, let us gather the four equations. Introducing the Mino time
\be
\label{E0}
\rd \tau = \frac{\rd \lambda}{\Sigma (r(\lambda), \theta(\lambda))} ,
\ee
the system is given by
\begin{subequations}
\label{elgeod_sys_m}
\begin{align}
\label{E1_m}
 m \frac{\rd t }{\rd \tau} &  = \frac{1}{\Delta} \left[ (r^2+a^2)^2 E - 2 a M r L \right]  - a^2 E \sin^2{\theta}  + a q \cP  \cos{\theta} \\
 \label{E2_m}
 m \frac{\rd r}{\rd \tau}  & = \epsilon_r \sqrt{R(r)}  \\
 \label{E3_m}
 m \frac{\rd \theta}{\rd \tau}  & = \epsilon_\theta \sqrt{\Theta(\theta)} \\
\label{E4_m}
 m \frac{\rd \varphi }{\rd \tau} & =  \frac{L + q\cP  \cos{\theta}}{\sin^2{\theta}} + \frac{a}{\Delta} \left(  2 M r E - a L \right) ,
\end{align}
\end{subequations}
where $\epsilon_r = \pm 1$, $\epsilon_\theta = \pm 1$ and
the radial and polar potentials are given by
\begin{align}
R(r) & = \left[ (r^2+a^2) E - a L\right]^2 - \Delta (r^2 m^2 + \cK)\\
\Theta(\theta) & = \cK - m^2 a^2 \cos^2{\theta} - \left( \frac{L + q\cP  \cos{\theta}}{\sin{\theta}} - a E \sin{\theta} \right)^2 .
\end{align}
As expected, this system of equations corresponds to the electrically neutral subsector of the equations derived and studied by Hackmann and Xu in \cite{Hackmann:2013pva}.

For later purposes, it will be useful to introduce the specific conserved energy $\varepsilon$, the specific conserved angular momentum $\ell$, the reduced Carter constant $k$
and the charge-to-mass ratio $\kappa$ by
\be \label{specific_quantities}
    \varepsilon = \frac{E}{m}, \quad 
    \ell = \frac{L}{m}, \quad
    k = \frac{\cK}{m^2}\quad \mbox{and} \quad
    \kappa = \frac{q}{m} .  
\ee
Note that, in the geometrical ($G=1$ and $c=1$) and Gaussian units that we are using,  
$\varepsilon$ and $\kappa$ are dimensionless, $\ell$ has the dimension of a length and $k$ that of a squared length. In terms of the reduced quantities \eqref{specific_quantities}, the system \eqref{elgeod_sys_m} becomes
\begin{subequations}
\label{elgeod_sys}
\begin{align}
\label{E1}
 \frac{\rd t }{\rd \tau} &  = \frac{1}{\Delta} \left[ (r^2+a^2)^2\varepsilon - 2 a M r \ell \right]  - a^2 \varepsilon \sin^2{\theta}  + a \kappa \cP  \cos{\theta} \\
 \label{E2}
 \frac{\rd r}{\rd \tau}  & = \epsilon_r \sqrt{\cR(r)}  \\
 \label{E3}
 \frac{\rd \theta}{\rd \tau}  & = \epsilon_\theta \sqrt{\tilde{\Theta}(\theta)} \\
\label{E4}
 \frac{\rd \varphi }{\rd \tau} & =  \frac{\ell + \kappa\cP  \cos{\theta}}{\sin^2{\theta}} + \frac{a}{\Delta} \left(  2 M r \varepsilon - a \ell \right) ,
\end{align}
\end{subequations}
with 
\begin{align}
\cR(r) & = \left[ (r^2+a^2) \varepsilon - a \ell\right]^2 - \Delta (r^2 + k)\\
\tilde{\Theta}(\theta) & = k - a^2 \cos^2{\theta} - \left( \frac{\ell + \kappa\cP  \cos{\theta}}{\sin{\theta}} - a \varepsilon \sin{\theta} \right)^2 .
\label{tTheta}
\end{align}
Note that $\cR(r) = R(r)/m^2$ and $\tilde{\Theta}(\theta) = \Theta(\theta)/m^2$ and the scaling have been chosen so that the particle's mass $m$ does not appear in the above formulas. 
Note as well that $\kappa\cP$ has the dimension of a length. 

The radial and polar equations (\ref{E2}) and (\ref{E3}) imply that
\begin{align}
\tau - \tau_0  & = \int^r_{r_0} \frac{\epsilon_r \rd \bar{r} }{\sqrt{\cR(\bar{r})}} = \int^{\theta}_{\theta_0} \frac{\epsilon_{\theta} \rd \bar{\theta} }{\sqrt{\tilde{\Theta}(\bar{\theta})}} .
\end{align}
Then, the time and azimuthal equations (\ref{E1}) and (\ref{E4}) can be integrated as 
\begin{align}
t -t_0 & = \int^{r}_{r_0} \frac{(\bar{r}^2 + a^2)^2 \varepsilon - 2 a M \bar{r} \ell}{\Delta(\bar{r})} \frac{\epsilon_r \rd \bar{r}}{\sqrt{\cR(\bar{r})}} -   \int^{\theta}_{\theta_0}  \left[ a^2 \varepsilon \sin^2{\bar{\theta}} - a \kappa \cP  \cos{\bar{\theta}}\right] \frac{\epsilon_{\theta} \rd \bar{\theta}}{\sqrt{\tilde{\Theta}(\bar{\theta})}} \\
\varphi - \varphi_0 & = a \int^r_{r_0} \frac{2 M \bar{r} \varepsilon - a \ell}{\Delta(\bar{r})} \frac{\epsilon_r \rd \bar{r}}{\sqrt{\cR(\bar{r})}} +   \int^{\theta}_{\theta_0} \left[ \frac{\ell + \kappa \cP  \cos{\bar{\theta}}}{ \sin^2{\bar{\theta}}} \right] \frac{\epsilon_{\theta} \rd \bar{\theta}}{\sqrt{\tilde{\Theta}(\bar{\theta})}} .
\end{align}
These integrals formulas provide a closed form solution for the electrogeodesic motion in the Kerr monopole magnetopshere. As expected, the correction terms proportional to $\kappa$ being purely polar dependent, they can be integrated out without further difficulty. 

Let us now point that the same conclusion is valid for the split monopole solution (\ref{spli}). Indeed, the separability which relies on the property (\ref{COND22}) remains valid even with the shift of the gauge potential in $\cP \rd \varphi$. The same system of separable equations for the electrogeodesics remains valid except that the angular momentum of the charged particle has to be shifted such as $L \rightarrow \tilde{L} = L - q\cP$. Therefore, the result we shall present in the following section also apply to the split monopole solution provided one takes into account this slight modification. We can now use this results to study the ejection of charged particles in this specific magnetosphere.

\section{Jet launching in the vacuum Kerr monopole}

\label{C}

In this section, we use the exact solution to the electrogeodesic equation derived in the previous section to study the acceleration of charged particles near the poles and the formation of jets. We present the exact formula for the 4-acceleration. Then, we classify the stable and unstable polar equilibrium positions. We further derive the expression for the azimuthal speed experienced by the charged particles in this jet, which reveals the role of the magnetic frame-dragging effect. Finally, we compute the electromagnetic gravitational redshift measured by an asymptotic Carter observer and determine the condition under which the ejected charged particle is blueshifted. To our knowledge, the only other analytical investigations of the motion of charged particles in the Kerr monopole have been discussed in \cite{Hackmann:2013pva} and more recently in \cite{Khan:2023ttq}.

We have seen that the electrogeodesic satisfies general conditions given by (\ref{cond1}), (\ref{cond2}) and (\ref{cond3}). Applied to the magnetic monopole magnetosphere given by
the 4-potential
[cf. Eq.~\eqref{tilde_A_alpha_0}]
\be
A = \frac{ \cP \cos{\theta}}{\Sigma} \left( a \rd t - (r^2+a^2) \rd \varphi\right) .
\ee
conditions (\ref{cond1}) and (\ref{cond3}) respectively give
\begin{align}
& \varepsilon > - \frac{a \kappa \cP \cos{\theta}}{\Sigma}  \quad\text{for}\quad r > r_{\rm ergo}(\theta)  \label{cond_eps_out_ergo} \\
& \varepsilon >  \Omega_{\rm C} \ell , \label{condgen}
\end{align}
where we have used property~\eqref{COND22} and have let appear the angular velocity $\Omega_{\rm C} = \Omega_{\rm C}(r)$ of the Carter observer, via $\Omega_{\rm C} = a/ (r^2 +a^2)$ [Eq.~\eqref{Omega_Carter}].
Along the Northern rotation axis ($\theta = 0$), where  $r_{\rm ergo}(0) = r_+$, we may rewrite \eqref{cond_eps_out_ergo} as
\begin{align}
\label{cond_rot_axis}
    \varepsilon > - \Omega_{\rm C} \kappa \cP  \qquad (\theta = 0) .
\end{align}
The above conditions will serve as consistency checks for the following discussion.

\subsection{Acceleration of charged particles}

With this solution at end, one can obtain an analytic formula for the 4-acceleration $a^\mu$ of the charged particles. The latter is given by the Lorentz force formula:
\be \label{a_kappa_F_u}
a^{\mu} = \kappa F^{\mu}{}_{\nu} u^{\nu} . 
\ee
It is easier to perform the computation in the Carter tetrad $(e_\alpha)$, as given by Eq.~\eqref{Carter_tetrad}.
The components of the particle's 4-velocity with respect to the coordinate frame are 
$u^\alpha = \Sigma^{-1} \rd x^\alpha / \rd \tau$ with $\rd x^\alpha / \rd \tau$ given by the electrogeodesic system \eqref{elgeod_sys}. Transforming to the Carter tetrad 
via the system \eqref{Carter_tetrad}, we get the following expression of the 4-velocity:
\begin{align}
    \label{4velocity_Carter_tetrad}
    u =  \frac{\varepsilon (r^2 + a^2) - a \ell}{\sqrt{\Sigma}\sqrt{\Delta}} \, e_0
        + \epsilon_r \frac{\sqrt{\cR(r)}}{\sqrt{\Sigma}\sqrt{\Delta}} \, e_1 
        + \epsilon_\theta \frac{\sqrt{\tilde{\Theta}(\theta)}}{\sqrt{\Sigma}} \, e_2  + \frac{\ell + \kappa \cP \cos\theta - a \varepsilon \sin^2\theta}{\sqrt{\Sigma}\sin\theta} \, e_3 .
\end{align}
Using the components \eqref{KYsol} of the Faraday tensor in the Carter tetrad, Eq.~\eqref{a_kappa_F_u} leads to 
the following components of the particle's 4-acceleration in the Carter frame\footnote{See Notebook~2 in App.~\ref{sage_notebooks} for the computation.}:
\begin{subequations} 
\label{acc}
\begin{align}
a^0 = & - \epsilon_r \frac{2 a \kappa \cP r \cos\theta \sqrt{\cR(r)}}{\Sigma^{5/2} \sqrt{\Delta}} \\
a^1 = & - \frac{2 a \kappa \cP  \left[ (r^2 + a^2) \varepsilon - a \ell \right]  r \cos\theta}{\Sigma^{5/2} \sqrt{\Delta} } \label{radialacc} \\
\label{polaracc}
a^2 = & \frac{\kappa  \cP (\ell + \kappa \cP \cos\theta - a \varepsilon \sin^2\theta) (r^2-a^2\cos^2\theta)}{\Sigma^{5/2} \sin\theta} \\
a^3 = & - \epsilon_\theta \frac{\kappa  \cP \sqrt{\tilde{\Theta}(\theta)}  (r^2-a^2\cos^2\theta)}{\Sigma^{5/2}} .
\end{align}
\end{subequations}
The norm of the 4-acceleration is found to be
\be
\label{norm}
    \sqrt{a_\mu a^\mu} = \frac{|\kappa \cP|}{\Sigma^{3/2}} \sqrt{ k + \frac{a^2}{\Sigma} (3 r^2 -  a^2 \cos^2\theta) \cos^2\theta } . 
\ee
Note that it involves only $|\kappa \cP|$ and the reduced Carter constant $k$. Let us now investigate what these analytic expressions teach us about the 4-acceleration experienced by the charged particles in the different regions.

Consider first the equatorial plane, i.e. $\theta =\pi/2$. The components $a^0$ and $a^1$ of the 4-acceleration are proportional to $\cos{\theta}$, so that they identically vanish at the equator:
\be
a^0 (r,\pi/2)= a^1 (r,\pi/2) = 0 .
\ee
As for the azimuthal and polar components, one finds that 
\begin{align}
a^2 (r,\pi/2) & =  \frac{\kappa  \cP (\ell  - a \varepsilon) }{r^{1/2} }  \\
a^3 (r,\pi/2) & = - \epsilon_\theta \kappa  \cP \frac{\sqrt{k - (\ell - a \varepsilon)^2}}{r^{1/2}} .
\end{align}
We see that for a general electrogeodesic labelled by $(\varepsilon, \ell, k)$, the polar acceleration will drive it away from the equatorial plane, suggesting that this position is no longer an equilibrium position as for the geodesic case. Yet, an exception remains: it corresponds to the electrogeodesics for which $\ell = a \varepsilon$
(note that this satisfies the inequality (\ref{condgen})). In this case, the polar acceleration vanishes, i.e. $a^2 (r,\pi/2) =0$, potentially allowing for an equilibrium position. This can be confirmed by the analysis of the polar potential which is the subject of the next subsection. 

On the contrary, near the poles, at $\theta =0$ (or equivalently $\theta =\pi$), the time and radial components of the acceleration $(a^0, a^1)$ take their maximal value, which suggests that charged particles are ejected preferentially near the poles. Notice that the time and radial components of the acceleration is directly proportional to (i) the spin $a$ of the black hole and (ii) the parameter $\kappa \cP$ which encodes the presence of the magnetosphere, demonstrating the combined roles of rotation and the magnetic monopole to radially accelerate charged particles. In this regard, this property of the radial acceleration of charged particles in the magnetic monopole Kerr magnetosphere parallels the behavior of the Poynting flux in the Blandford-Znajek mechanism, which is also triggered by both the rotation of the black hole and the magnetic monopole charge of the magnetosphere.
Finally, let us also point that thanks to condition (\ref{condgen}), the term $(r^2 + a^2) \varepsilon - a \ell$ in the radial acceleration \eqref{radialacc} cannot vanish. 

There is a location at which the particle's 4-acceleration $a^\mu$ vanishes. Since $a^\mu$ is necessarily a spacelike vector, its vanishing is equivalent to its norm $\sqrt{a_\mu a^\mu}$ being zero. From expression \eqref{norm} of the latter, we get the condition 
\be
\label{zeroacc}
r = r_{\ast}(\theta) = a |\cos{\theta}| \sqrt{\frac{a^2 \cos^2\theta - k}{3 a^2\cos^2\theta + k}} .
\ee
At $r = r_{\ast}(\theta)$, the electrogeodesics reduces to a free-fall motion. However,  since the Carter constant has to satisfy $ 0 < k < a^2 \cos^2{\theta}$, this makes this hypersurface unphysical. To see this, let us analyze the locus of points for each family of electrogeodesic satisfying either $k=0$ or $k = a^2\cos^2{\theta}$. It follows that for each family, one has
\begin{align}
r(\theta) |_{k=0} = \frac{|\cos{\theta}|}{\sqrt{3}} a \leqslant \frac{a}{\sqrt{3}} \leqslant \frac{M}{\sqrt{3}}  < M \qquad r(\theta) |_{k=a^2 \cos^2{\theta}} = 0
\end{align}
so that none of the values allowed for $k$ and $a$ allows this hypersurface to be outside the outer horizon.

Finally, it is interesting to obtain the behavior of the acceleration at large $r$; it reads
\begin{subequations}
\begin{align}
a^0 \sim a^{(t)} \sim  & \; - 2 \epsilon_r a \kappa \cP \sqrt{\varepsilon^2 - 1} \, \frac{\cos\theta}{r^3} \\
a^1 \sim a^{(r)} \sim & \; - 2 a \kappa \cP \varepsilon \, \frac{\cos\theta}{r^3} \\
a^2 \sim a^{(\theta)} \sim & \; \frac{\kappa  \cP (\ell + \kappa \cP \cos\theta - a \varepsilon \sin^2\theta)}{r^3 \sin\theta} \\
a^3 \sim a^{(\varphi)} \sim & \; - \epsilon_\theta \frac{\kappa \cP \sqrt{\tilde{\Theta}(\theta)}}{r^3} ,
\end{align}
\end{subequations}
while the norm behaves as 
\be
     \sqrt{a_\mu a^\mu} \sim \frac{|\kappa \cP|}{r^3}  \sqrt{k + 3 a^2 \cos^2\theta} .
\ee
This concludes our discussion on the acceleration. It is crucial to notice that the present analysis has been made possible thanks to the exact solution of the electrogeodesic motion. As we can see, the expression for the acceleration experienced by the test charged particles already reveals several interesting features of the magnetosphere and the motion it induces on the particles. The key point is that even in this vacuum magnetosphere, the charged particles are naturally driven towards the poles and preferentially ejected near the poles, the ejection (i.e. the radial acceleration) being triggered both by the rotation and the magnetic monopole carried by the black hole. 

\subsection{Structure of the jet}

We now study how to characterize the jet. To this end, it is useful to identify the equilibrium positions of the polar motion and the stability of these positions. Since the aligned jet corresponds to charged particles escaping at infinity along the rotation axis, we first focus on characterizing the conditions for the rotation axis to be a stable position. As we shall see, the presence of the magnetosphere introduces two regimes depending on the energy of the charged particles. Then, we further discuss the case of the equatorial plane. 

\subsubsection{Stable versus unstable polar equilibrium positions}

Consider the charged particles which can cross the rotation axis located $\theta=0$ in the Northern hemisphere.
From expression~\eqref{tTheta} of $\tilde\Theta(\theta)$, this requires 
\beq \label{ell_m_kap_P}
    \ell = - \kappa \cP 
\eeq
in order to cancel the divergence in the potential. Notice that in the neutral case, one recovers the standard result that $\ell=0$ is required to reach the rotation axis. Moreover, 
for the value \eqref{ell_m_kap_P} of $\ell$, the general condition (\ref{condgen}), i.e. $\varepsilon > \Omega_{\rm C} \ell$, reduces to 
condition \eqref{cond_rot_axis}, namely $\varepsilon > - \Omega_{\rm C} \kappa \cP$.

\begin{figure}
   \centerline{ \includegraphics[width=0.48\textwidth]{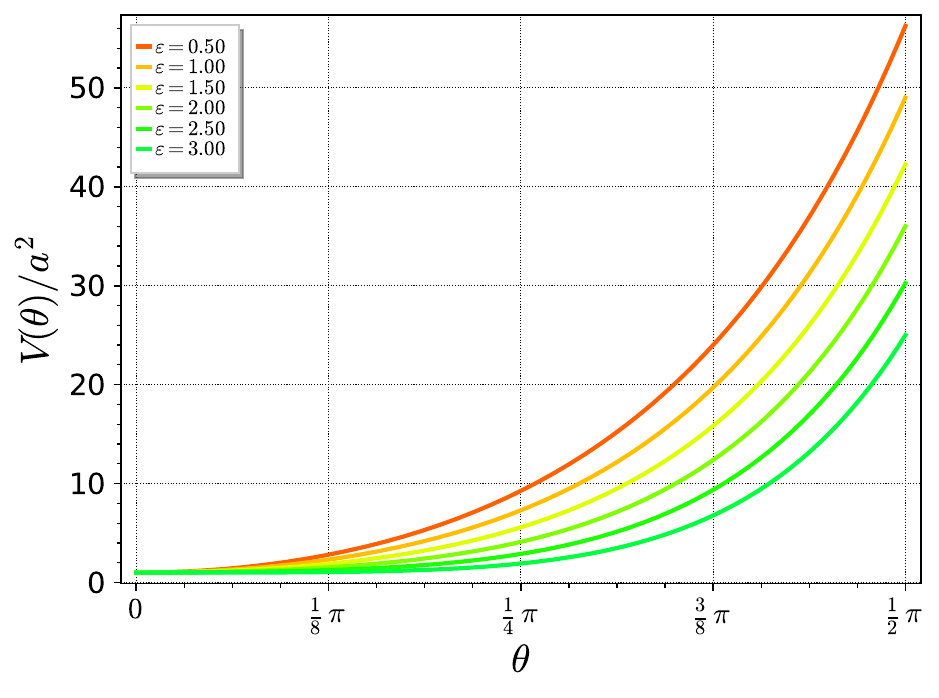}
    \includegraphics[width=0.48\textwidth]{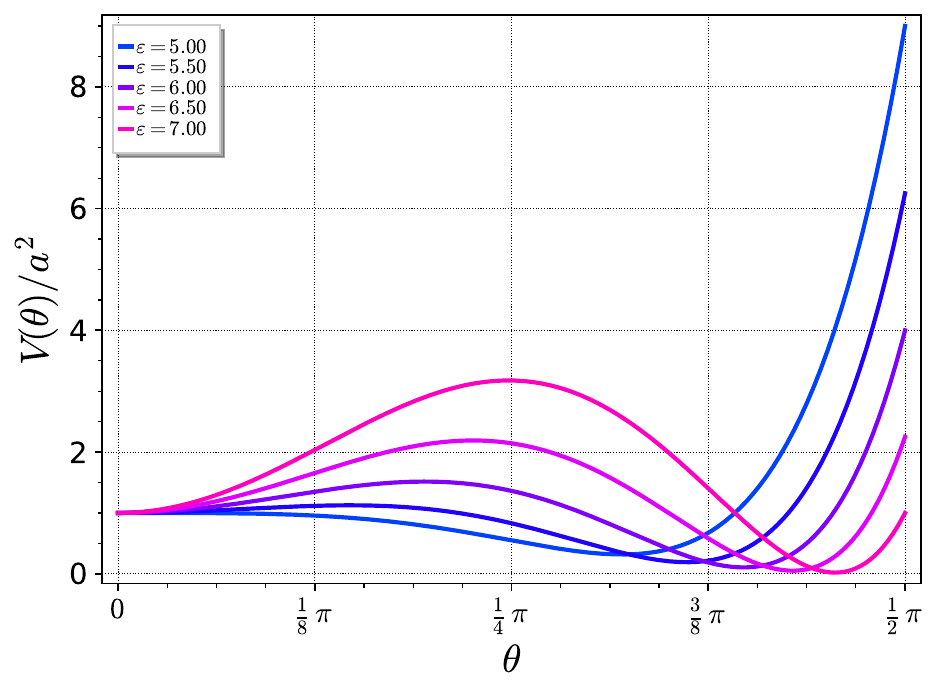} }
\caption{\label{pot_stable_theta_0} Effective potential $V(\theta)$ governing the latitudinal motion for $\ell=  - \kappa \cP$. The plots are performed 
for $\beta = 8$ and 
for various values of $\varepsilon$ ensuring a stable equilibrium at $\theta = 0$, according to Eq.~\eqref{theta_0_stable}, i.e.
$\varepsilon < 3$ (left panel) or $\varepsilon > 5$ (right panel). Notice that in all cases, $\theta=0$ is a (local) minimum of $V$
and that the marginal cases $\varepsilon=3$ and $\varepsilon=5$ are included in the plots. 
}
\end{figure}

\begin{figure}
   \centerline{ \includegraphics[width=0.48\textwidth]{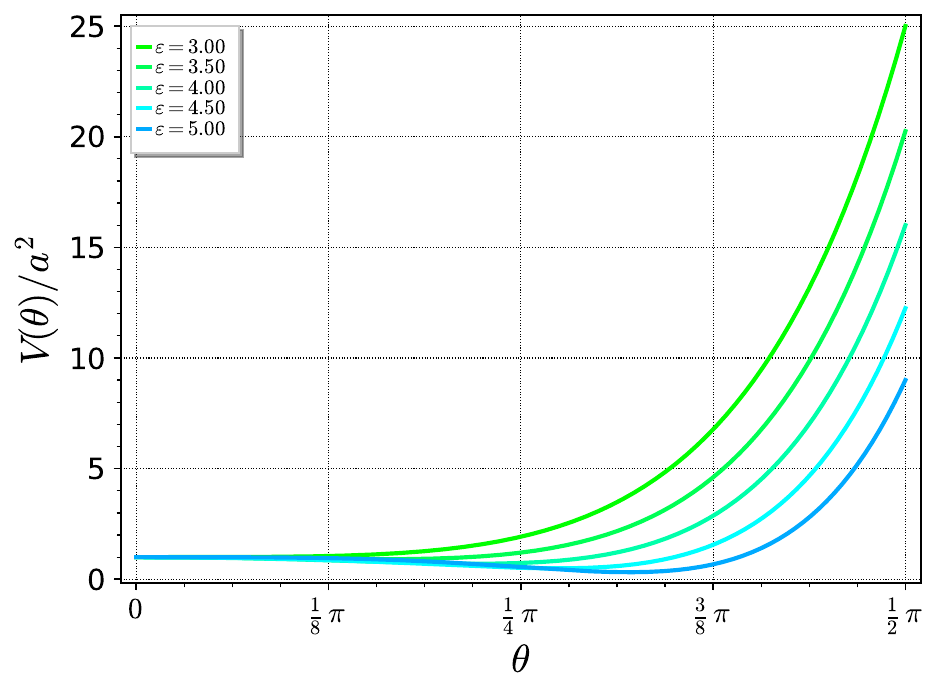} }
\caption{\label{pot_unstable_theta_0} Effective potential $V(\theta)$ governing the latitudinal motion for $\ell=  - \kappa \cP$ and $\beta=8$.
Contrary to Fig.~\ref{pot_stable_theta_0}, the chosen values of $\varepsilon$ are such that the equilibrium position at $\theta = 0$ is unstable:
it corresponds to a local maximum of $V(\theta)$. According to Eq.~\eqref{theta_0_stable} with $\beta=8$, these values are in the range
$3 < \varepsilon < 5$ (the marginal cases $\varepsilon=3$ and $\varepsilon=5$ are included in the plots). 
Notice that a stable equilibrium position occurs at $\theta = \theta_* \in (0,\pi/2)$.}
\end{figure}

Sticking to the value \eqref{ell_m_kap_P} of $\ell$ and assuming $a\neq 0$, the latitudinal equation of motion \eqref{E3} can be rewritten as
\beq
    \left( \frac{\rd \theta}{\rd \tau} \right) ^2 + V(\theta) = k 
\eeq
with the effective potential\footnote{Use has been made of the identity $(1-\cos\theta)/\sin\theta = \tan(\theta/2)$.}
\beq
    V(\theta) & = a^2 \left[ \cos^2\theta + \left( \varepsilon \sin\theta - \beta \tan \frac{\theta}{2} \right) ^2 \right] 
\eeq
where $\beta$ is the dimensionless quantity defined by 
\beq \label{def_beta}
    \beta = - \frac{\kappa\cP}{a} = \frac{\ell}{a} .
\eeq
As we shall see, this parameter is crucial to characterize the new features induced by the magnetosphere. In term of $\beta$, condition (\ref{condgen}) reads 
\be \label{condgen_beta}
\varepsilon > a \Omega_C \beta . 
\ee
By definition, an equilibrium latitudinal position for the charged particles satisfies
\be
V(\theta) = k, \qquad V'(\theta) =0
\ee
and it is stable if $V''(\theta) > 0$ and unstable if $V''(\theta) < 0$. The effective potential $V(\theta)$ is depicted in Figs.~\ref{pot_stable_theta_0}--\ref{pot_unstable_theta_0}. 

Let us first consider the North pole $\theta =0$, which corresponds to the aligned jet, i.e. to charged particles escaping to infinity along the rotation axis. In this case, one finds 
\begin{align}
 V(0) & = a^2,\qquad V'(0)  = 0,\qquad V''(0)  = 2a^2 ( \varepsilon - \beta/2 - 1 )( \varepsilon - \beta/2 + 1) .
 \end{align}
It follows that $\theta=0$ is a stable constant latitudinal position, i.e. that $\rd \theta/\rd \tau =0$ and $V''(0)>0$, if and only if, the electrogeodesic $(\varepsilon, \ell, k)$ satisfies 
\begin{align}
\label{theta_0_stable}
    k & = a^2 \qquad   \text{and} \qquad  \varepsilon <  \frac{\beta}{2} - 1  \quad \mbox{or} \quad   \varepsilon > 1 + \frac{\beta}{2} .
\end{align}
Taking into account condition \eqref{condgen_beta}, one obtains the following conditions:
\begin{align}
\label{condpol}
 a \Omega_{\rm C} \beta < \varepsilon <  \frac{\beta}{2}-1   \qquad \mbox{or} \qquad  \varepsilon > \frac{\beta}{2} +1 .
\end{align}
For the second condition, we have taken into account that $a\Omega_{\rm C} = a^2 / (r^2 + a^2)$ always fulfills $a\Omega_{\rm C} < 1/2$, with the upper bound $1/2$ approached only near the event horizon of
an extremal ($a=M$) Kerr black hole, so that $\varepsilon > a \Omega_C \beta$ is actually implied by $\varepsilon > {\beta}/{2} +1$. It follows that there are two distinct families of electrogeodesics defined by the above thresholds, which correspond to such and aligned jet. 

Beside $\theta=0$, other possible latitudinal equilibrium positions can be found from the zeros of $V'(\theta)$, which reads, for  $\ell = - \kappa \cP$, 
\begin{align}
 V'(\theta) = \frac{2a^2}{\sin^2\theta} \left\{ 2 \beta^2 \tan \frac{\theta}{2}  + \sin\theta \left[ \left( (\varepsilon^2 - 1) \cos\theta - \varepsilon \beta \right) \sin^2\theta 
    - \beta^2 \right]\right\} .
\end{align}
Figures~\ref{pot_stable_theta_0} (right panel) and \ref{pot_unstable_theta_0} reveal that stable latitudinal equilibria at $\theta = \theta_* \neq 0$ exist for some values of the parameters. They are also of interest as they may correspond to misaligned jets of charged particles. Their characterization is however more involved and shall be discussed in a future work. 

To conclude this section, let us comment on the parameter $\beta$. First, notice that for an electron\footnote{In terms of SI units, one has actually $\kappa = \sqrt{\mu_0/(4\pi)}\,  q/m$, so that, for an electron, 
$\kappa_{\rm e}=-5.56\; 10^{7} \; {\rm m}^{1/2} {\rm kg}^{-1/2}$, which yields $\kappa_{\rm e} = -2.04 \; 10^{21}$ in geometrical units.}, $\kappa = \kappa_{\rm e} = -2.04 \; 10^{21}$,
while for a proton, $\kappa = \kappa_{\rm p} = 1.11 \; 10^{18}$. These huge values reflect the large dominance of the electromagnetic interaction over the gravitational one for elementary particles.  
On the contrary, $\cP$ must obey $|\cP| \ll M$, in order for 
the electromagnetic field to not significantly alter the spacetime metric, which is assumed to be the Kerr metric; otherwise one would have to 
consider the Kerr-Newman metric. For values of $a \sim M$, this is equivalent to demand $|\cP|/a \ll 1$. It follows that $\beta$ can take
values of order one or even much higher. For instance, for electrons, for which $|\kappa| \sim 10^{21}$, $\beta$ can be as large as $10^{10}$ for $|\cP| = 10^{-10} M$. This confirms that the model can account for large magnetization without breaking the test field assumption.

\subsubsection{Azimuthal speed and magnetic frame-dragging}

Let us now discuss the azimuthal angular velocity experienced by the charged particles. Imposing $\ell= - \kappa \cP$, one obtains
\begin{align}
 \frac{\rd t }{\rd \tau} &  = \frac{1}{\Delta} \left[ (r^2+a^2)^2\varepsilon + 2 \kappa a M \cP  r  \right]  - a^2 \varepsilon \sin^2{\theta}  + a \kappa \cP  \cos{\theta} \\
 \frac{\rd \varphi }{\rd \tau} & =  \frac{a}{\Delta} \left(  2 M r \varepsilon + a \kappa \cP \right) +  \frac{ \kappa\cP \left(  \cos{\theta} -1 \right)}{\sin^2{\theta}} .
\end{align}
The angular velocity is then given by
\be
\label{MFD}
\Omega (r,\theta) = \frac{\rd \varphi}{\rd t} = \frac{\left( 2Mr \varepsilon + a \kappa \cP\right) a \sin^2{\theta} + \kappa \cP \left(  \cos{\theta} -1 \right) \Delta }{\left[ (r^2+a^2)^2 \varepsilon + 2 \kappa a M \cP r + a\Delta \left(\kappa \cP \cos{\theta} - a \varepsilon \sin^2{\theta} \right)\right] \sin^2{\theta}} .
\ee
As expected, this angular velocity is not defined for particles located at the poles, i.e. at $\theta =0$. However, this formula provides the angular velocity of the particle for $\theta \in \; ] 0,\pi/2]$. In particular, if the jet is misaligned with the black hole rotation axis, i.e. $\theta\neq0$, this angular velocity encodes the precession of the jet of charged particles. Such precession of a misaligned jet has been recently measured for the M87* supermassive black hole and reported in \cite{Cui:2023uyb}.

Let us now evaluate the expression of $\Omega (r,\theta)$ in the large $r$ limit. Before proceeding, notice that in the absence of a magnetosphere, i.e. $\cP=0$, one recovers the standard gravitational frame-dragging effect of the Kerr black hole:
\be
\Omega^{\text{K}}_{\infty} (r, \theta) \sim \frac{2 a M}{r^3} .
\ee
The magnetic monopole dramatically changes this effect. When the magnetosphere is turned on, the large distance behavior is dominated by the electromagnetic interaction: the azimuthal speed experienced by charged particle in our magnetized Kerr (MK) becomes
\be
\label{angspeed}
\Omega^{\text{MK}}_{\infty} (r, \theta) \sim -  \frac{\kappa \cP}{\varepsilon r^2} \frac{(1- \cos{\theta} )}{ \sin^2{\theta}} .
\ee
We observe that for $\cP >0$, the angular velocity has a sign opposite to that of $\kappa$, i.e.  the magnetosphere forces positive charges to have a retrograde motion, while negative charges have a prograde motion. 
For a collimated jet, i.e. for a small opening angle $\theta_{\rm m}$, it reduces to 
\be
\Omega^{\text{MK}}_{\infty} (r, \theta_{\rm m}) \sim  - \frac{ \kappa \cP}{2\varepsilon r^2} \left( 1 + \frac{\theta_{\rm m}^2}{4} \right) + \cO(\theta_{\rm m}^4) . 
\ee
Remarkably, it depends solely on the magnetic monopole charge $\cP$, the charge-to-mass ratio $\kappa$ and the specific energy of the particle $\varepsilon$. 
At the equator $\theta =\pi/2$, Eq.~\eqref{angspeed} yields
\be
\Omega^{\text{MK}}_{\infty} (r, \theta) \sim -\frac{\kappa \cP}{\varepsilon r^2} .
\ee
To conclude, the magnetosphere induces a new magnetic frame-dragging effect which completely dominates the standard gravitational frame dragging for electrogeodesics with $\ell= - \kappa \cP$. Moreover, one notices that while the gravitational frame dragging effect decays as $1/r^3$ at large distance, the magnetic frame dragging decays only as $1/r^2$, showing that it remains effective on much larger distances away from the black hole.

\subsection{Blueshift of charged particles and  the zone of acceleration }

We are now interested in characterizing what a Carter observer would measure when a charged particle crosses its trajectory.  For that purpose, it will reveal useful to introduce the 
redshift $z$ by 
\be \label{def_redshift}
 1 + z = \frac{E_{\text{em}}}{E_{\text{obs}}} , 
\ee 
where $E_{\text{em}}$ is the energy of the particle when emitted and $E_{\text{obs}}$ the energy of the particle when observed, both energies being measured by the Carter observer.

\begin{figure}
    \begin{center}
    \includegraphics[width=0.48\textwidth]{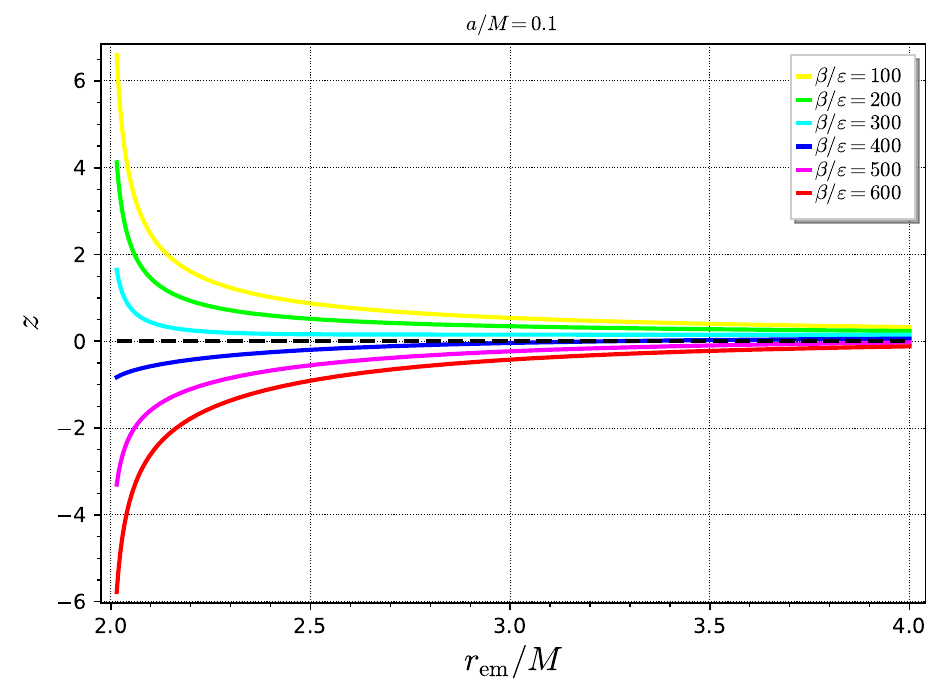}
    \includegraphics[width=0.48\textwidth]{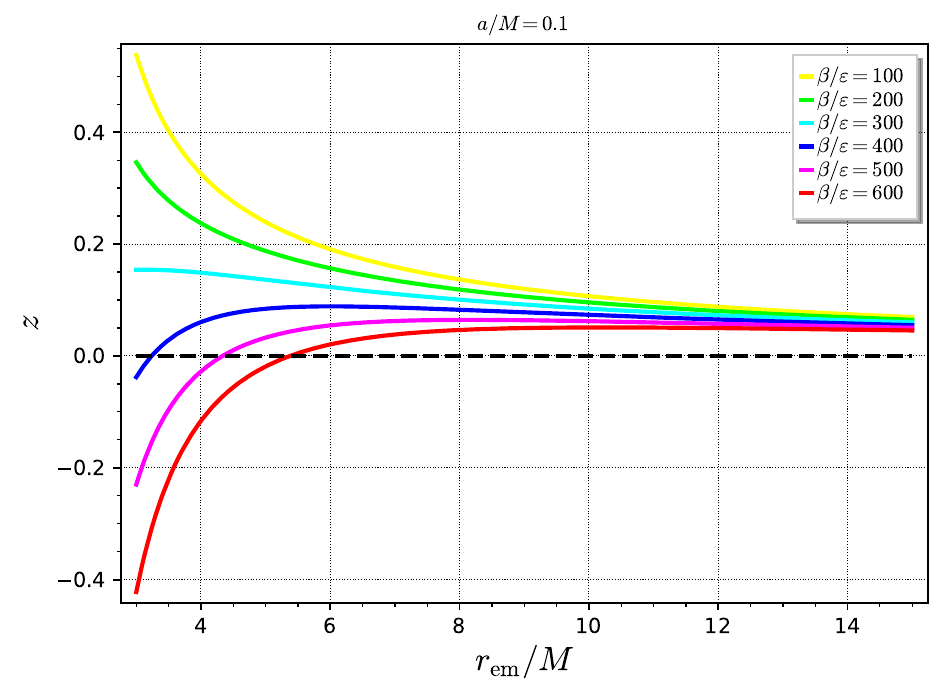} \\
    \includegraphics[width=0.48\textwidth]{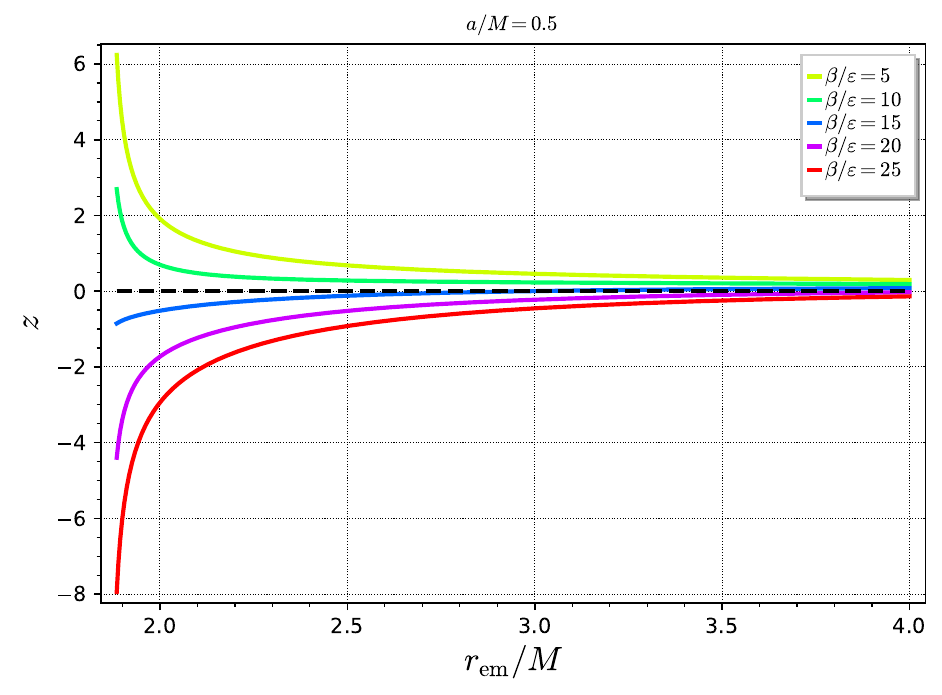}
    \includegraphics[width=0.48\textwidth]{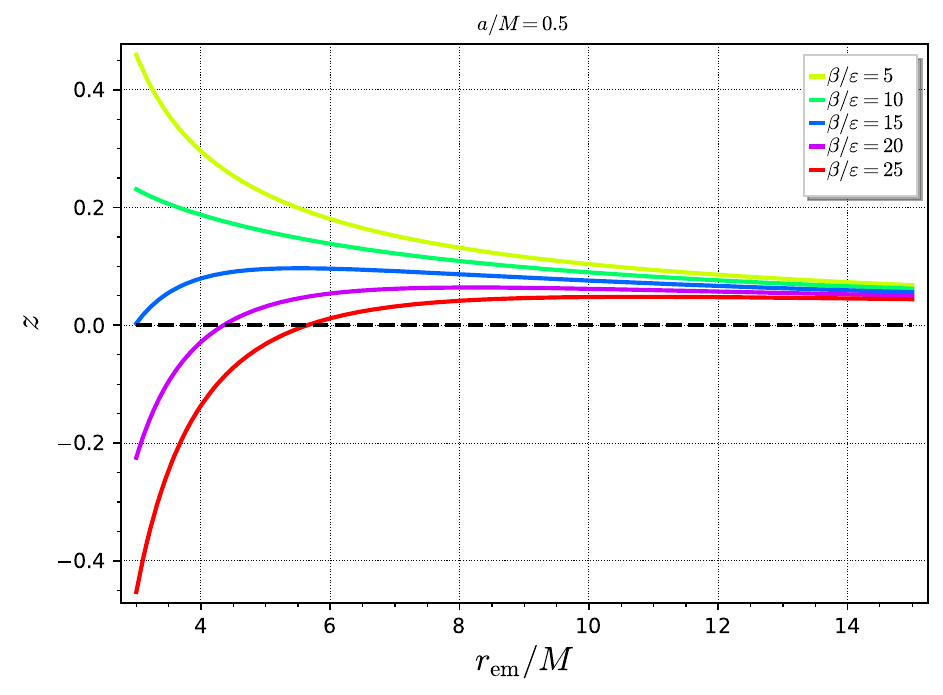} 
    \end{center}
\caption{\label{z_r_em_010_050} Redshift factor $z$ along the Northern rotation axis ($\theta=0$) as a function of the emission radius $r_{\rm em}$, for
various values of the ratio $\beta/\varepsilon$ and for $a/M=0.1$ (top figures) and $a/M=0.5$ (bottom figures).
Since $z$ is diverging for $r \to r_+$ (the horizon), the plots in the left figures start at $r = 1.01 \, r_+$. 
}
\end{figure}

\begin{figure}
    \begin{center}
    \includegraphics[width=0.48\textwidth]{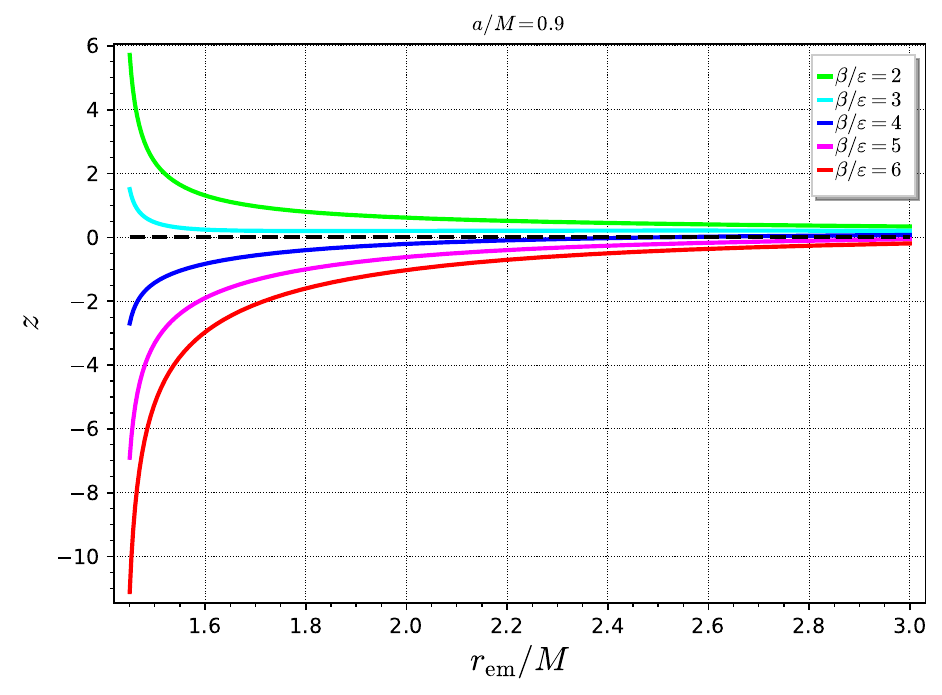}
    \includegraphics[width=0.48\textwidth]{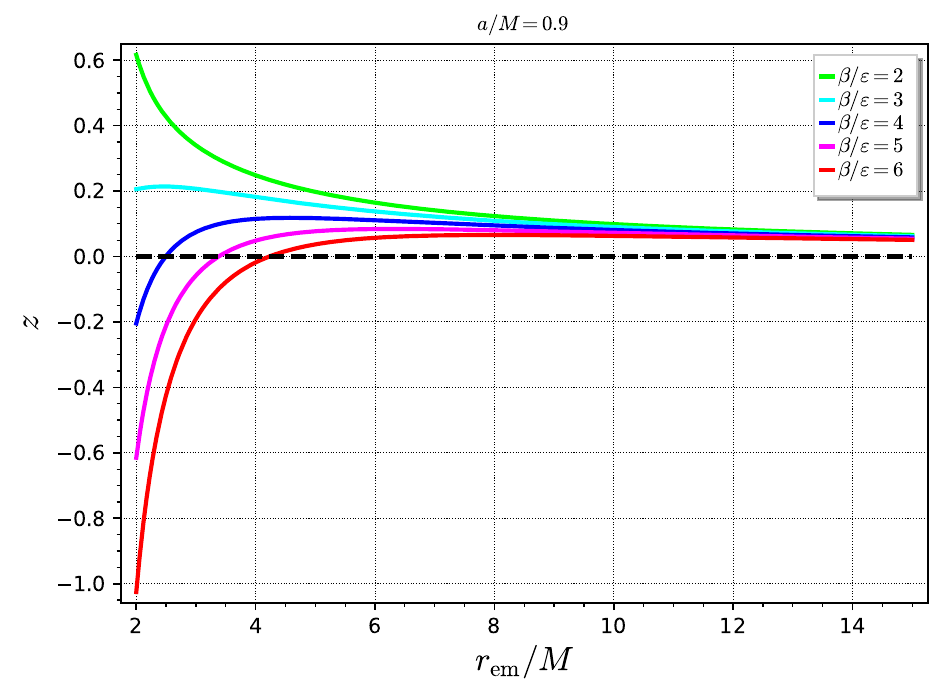} \\
    \includegraphics[width=0.48\textwidth]{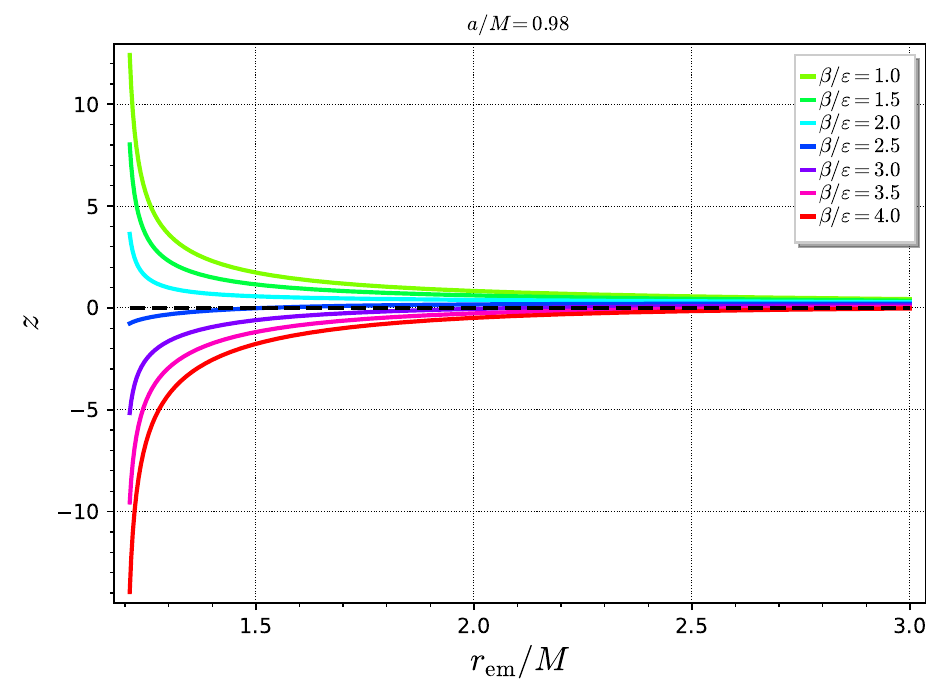}
    \includegraphics[width=0.48\textwidth]{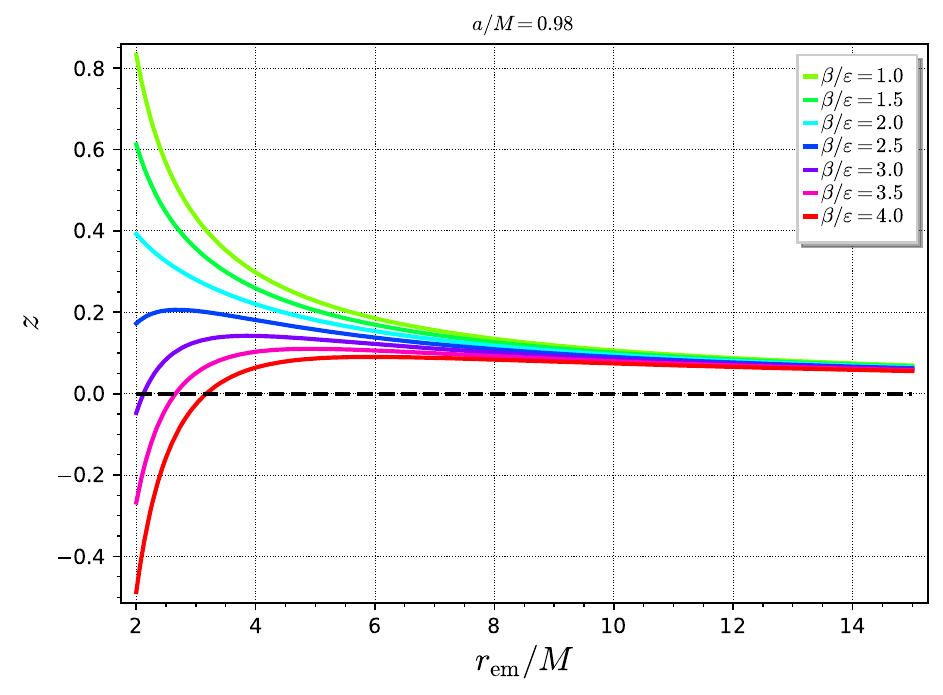} 
    \end{center}
\caption{\label{z_r_em_090_098} Same as Fig.~\ref{z_r_em_010_050} but for $a/M=0.9$ (top figures) and $a/M=0.98$ (bottom figures).
}
\end{figure}

The particle is blueshifted when $E_{\text{obs}} > E_{\text{em}}$ which corresponds to $z <0$.   Let us denote $r_{\text{em}}$  the radius at which the particle is emitted and let us consider that the particle is observed at $r \rightarrow +\infty$. We recall that the Carter observer has a $4$-velocity given by $e_0$ [cf. Eq.~\eqref{e_0_BL}].
The redshift factor \eqref{def_redshift} can thus be written as 
\begin{align}
1+ z = \frac{[u^{\mu}( e_0)_{\mu}]_{\text{em}}}{[u^{\mu}( e_0)_{\mu}]_{\infty}} .
\end{align}
We read on the components \eqref{4velocity_Carter_tetrad} of $u$ with respect to the Carter tetrad that 
\be
\label{enC}
u^{\mu}( e_0)_{\mu}  =  \frac{\varepsilon (r^2 + a^2) - a \ell}{\sqrt{\Sigma \Delta}} .
\ee
This formula yields $[u^{\mu}( e_0)_{\mu}]_{\infty} = \varepsilon$. Hence we recover that the conserved quantity $\varepsilon$ is interpretable as the particle's specific energy measured by an inertial
observer at infinity and at rest with respect to the black hole (such as the asymptotic Carter observer). 
For the case $\ell = -\kappa \cP$  that we are considering [cf. Eq.~\eqref{ell_m_kap_P}], we may write 
$\ell =  a \beta$ [cf. Eq.~\eqref{def_beta}] and a direct computation then gives 
\begin{align}
\label{redshift}
z = \frac{1}{\sqrt{\Sigma \Delta}} \left[  r_{\text{em}}^2 + a^2 \left(  1- \frac{ \beta}{\varepsilon} \right) -\sqrt{(r_{\text{em}}^2+a^2\cos^2{\theta}) (r_{\text{em}}^2 - 2M r_{\text{em}} + a^2 )} \right] . 
\end{align}
The particle is blueshifted when the numerator is negative. This happens for sufficiently large values of $\beta/\varepsilon = -\kappa \cP / (a \varepsilon)$ and $r_{\text{em}}$ sufficiently small, 
as shown in Figs.~\ref{z_r_em_010_050}--\ref{z_r_em_090_098}.

First, fixing the black hole spin $a/M$, we observe that the extent of the region from which ejected particles appear blueshifted at infinity, i.e. have $z<0$, grows with the ratio $\beta/\varepsilon$. For a slowly rotating black hole, e.g. $a = 0.1 M$ (top of Fig.~\ref{z_r_em_010_050}), ejected particles are blueshifted only for very high values of the parameter $\beta/\varepsilon$. For higher value of the spin (bottom of Fig.~\ref{z_r_em_010_050} and Fig.~\ref{z_r_em_090_098}), the required values of the parameter $\beta/\varepsilon$ for having a region of blueshift are much smaller. This behavior demonstrates the combined roles of both the spin and the magnetization of the black hole in accelerating charged particles for asymptotic observers.

Moreover, we observe on Figs.~\ref{z_r_em_010_050}--\ref{z_r_em_090_098} that the region in which emitted particles are blueshifted admits a maximal radius, which corresponds to $z=0$. This maximal radius is always close to the horizon. For instance, for $a=0.1M$ and $\beta/\varepsilon =600$ or $a=0.5M$ and $\beta/\varepsilon =25$, the maximal radius is at $r_{\ast} \sim 5.8 M > r_{+}$. The existence of this bounded region allowing charged particles to be blueshifted at infinity and its analytical description stand as the main results of this work. Within this specific model, this shows that accelerated particles of a given energy can only emerge from a finite region surrounding the black hole horizon.

To conclude, let us contrast our results with earlier investigations exploring the spontaneous collimation of charges particles near the charged Kerr black hole \cite{deFelice:1999ch}. In this work, the authors considered vortical electrogeodesics, i.e. trajectories which never encounter the equator. The Maxwell solution is chosen to be the purely electric sector of the Kerr-Newman black hole in the test field approximation. Focusing on these vortical electrogeodesics which correspond to $K < (L-aE)^2$, the authors numerically computed the associated Lorentz factor for those ejected charged particles. In the present work, we have chosen a different path and reached different results: the Maxwell solution corresponds to the magnetic sector, which is better motivated than the electric one and which can be easily adapted (by cutting the solution at the equator and flipping orientation of the Maxwell field in the other hemisphere) to the standard split-monopole solution (\ref{spli}). Second, we have chosen to focus on a different family of electrogeodesics, ie.e. the ones which can reach the rotation axis. Those choices have allowed us to provide a fully analytic solution for the ejected particles and their associated Lorentz factors, providing a major improvement to these earlier results.




\section{Discussion}

\label{D}

In this work, we have studied the ejection of charged particles around a magnetized Kerr black hole and the build-up of a relativistic jet near the poles. Contrary to most approaches, our investigation is fully analytic and provides close expressions for key quantities characterizing the jet (acceleration of the charged particles, polar equilibrium, frame-dragging effect, blueshift of the particle energy). Let us briefly summarize the main points of this analytic model.

The model relies on the simplest Kerr magnetosphere, which consists of the vacuum magnetic monopole configuration (\ref{tilde_A_alpha_0}) and naturally extends to the more physical split-monopole solution (\ref{spli}) with an appropriate shift of the angular momentum of the charged particles. As we reviewed in  Sec.~\ref{A}, these exact solutions are selected from the fundamental symmetries of the Kerr geometry and corresponds to the Penrose current (\ref{Penrose}) of the Kerr black hole. Therefore, it is intimately tied to the existence of a Killing-Yano tensor in this geometry. In turn, this exact solution of the source-free Maxwell equations in the Kerr geometry can also be recovered as the test field limit of the Kerr-Newman magnetosphere with a non-zero magnetic charge and a zero electric one. The key motivation for focusing on this Kerr magnetosphere model is that it belongs to the few exact solutions for which the motion of charged particles is fully separable. In Sec.~\ref{B}, we have reviewed in detail the geometrical condition (\ref{COND}) for a magnetosphere to preserve the Carter constant which ultimately ensures the separability, and have shown that among the models of magnetosphere built from the Killing symmetries, only the Killing-Maxwell system, based on the principal tensor, and the Penrose current, based on the KY tensor, satisfy this key condition. The first two sections thus provide a general self-contained review on (i) the different models of Kerr magnetospheres based on the Killing symmetries and (ii) the necessary conditions for the electrogeodesic motion to be separable. While most of the results presented in these first two sections are known, although not in this self-contained form, we have also included new results. In particular, in Sec.~\ref{A3}, we have shown that besides the Wald construction, one can also use the two commuting Killing vectors and the Hodge duality to build a new non-vacuum exact solution of the test Maxwell equations. This provides the new solution (\ref{newsol}), which describes a radially growing toroidal magnetic field surrounding the Kerr geometry. Whether this new solution can be useful in the vicinity of the black hole remains to be studied. This concludes the summary of the two sections intended to review the required material for our investigation of the motion of charged particles in the magnetized Kerr black hole (Secs.~\ref{secA} and \ref{B}). Let us now summarize the application to the jet launching mechanism.

Using the analytic solution of the electrogeodesic equation in the Kerr monopole vacuum magnetosphere, we have studied several key properties of the jet. 
\begin{itemize}
\item \textit{The form of the acceleration:} We have presented the analytic expression for the $4$-velocity $u^{\mu}$ [Eq.~(\ref{4velocity_Carter_tetrad})] and 4-acceleration $a^{\mu}$ [Eq.~(\ref{acc})] of the charged particles in the entire black hole exterior, which allowed us to contemplate several key properties of the system. Contrary to the azimuthal and polar acceleration $(a^{\theta}, a^{\varphi}$), the radial acceleration $a^r$ is shown to be proportional to both the magnetic charge $\cP$ (carried by the magnetosphere) and the black hole spin $a$, so that it vanishes in the non-rotating and/or non-magnetized case. This demonstrates the key role played by the presence of a spinning magnetically charged magnetosphere in the ejection mechanism. Second, the polar acceleration turns out to be maximal at the equator and vanishing at the pole. In contrast, the radial acceleration vanishes at the equator and becomes maximal at the poles, suggesting the build up of a jet of charged particles preferentially around the poles. For $\cP>0$ and negatively charged particles, $\kappa <0$, the acceleration is outwards and the particles are ejected from the North pole.
\item \textit{Polar equilibrium positions: } We have analyzed the stable and unstable equilibrium positions of the polar motion depending on the energy and angular momentum $(\varepsilon, \ell)$ of the charged particle.  The difference with the geodesic motion is best parametrized by a new parameter $\beta = -\kappa \cP /a$, where $\kappa$ is the charge-to-mass ratio of the particle.  First, particles can reach the rotation axis $\theta=0$ only for $\ell = a\beta$. Focusing on this family of electrogeodesics, we have identified a new condition, (\ref{theta_0_stable}), for the (North) rotation axis to be a stable polar position: the energy of the particle 
must be either smaller or greater than a threshold depending on $\beta$, cf. Eq~(\ref{condpol}). This case corresponds to an idealized aligned jet. Instead, trajectories violating this condition can admit a stable polar position at some $\theta_{\ast} \in \; ] 0, \pi/2[$ and thus correspond to a misaligned jet of particles. 
\item \textit{Magnetic frame-dragging effect:} We have studied the fate of the frame-dragging effect in the Kerr magnetic monopole. Using the expression of the azimuthal angular velocity for $\ell= - \kappa \cP$, i.e. Eq.~(\ref{MFD}), we have shown that the magnetic monopole introduces a new frame-dragging, which completely dominates the gravitational frame-dragging. Indeed, the former decays as $1/r^2$ while the latter decays as $1/r^3$, showing that the magnetic frame-dragging impacts charged particles on larger scales, even in this test field approximation. Furthermore, in contrast to the gravitational frame-dragging, which forces particles to rotate in a prograde trajectory, the magnetic counterpart forces particles to have prograde or retrograde motion depending on their electric charge. We emphasize that for misaligned jets of particles, this angular velocity shall encode the precession of the charged particles, and thus of the jet, around the rotation axis of the black hole. It would be interesting to confront this new formula to the recently measured precession of the M87* jet \cite{Cui:2023uyb} and explore to which extent this could allow one to constrain the mass and spin of the black hole. 
\item \textit{Geometry of the acceleration zone and asymptotic blueshift of charged particles}: We have further characterized the  condition under which charged particles ejected from the magnetosphere reach as asymptotic observer with a higher energy. By evaluating the redshift $z$ experienced by the particle [Eq.~(\ref{redshift})] we have provided an analytical condition, i.e. $z>0$, for the particle to be blueshifted at large distances. Interestingly, this condition reveals the existence of a structure, through a radius which encodes the maximal extent of the region in which emitted particles escaping to infinity will be observed as blueshifted. The maximal radius of this region grows with the spin and the magnetization of the black hole. 
\item \textit{Numerical values for the parameters:}  Our model is self-consistent for concrete realistic numerical values of the parameters. Indeed, as we have worked in the test field approximation, $\cP$ must obey $|\cP| \ll M$. We have seen that for the values of the charge-to-mass ratio $\kappa$ of the electron or proton, and for a spin of order $a \sim M$, the parameter $\beta$ can take
values of order one or even much higher while still ensuring that $|\cP| \ll M$. This confirms that the model can account for large magnetization without breaking the test field assumption. 
\end{itemize}
To our knowledge, the present model provides the simplest analytical study for the ejection of charged particles from the Kerr black hole magnetosphere, with a direct application to the split-monopole configuration (\ref{spli}). It offers a first minimal model to study the jet launching within a source-free magnetosphere based solely on the electrogeodesic motion, thus addressing one of the key challenges mentioned in the introduction. There are several open interesting directions suggested by our results. 

The present investigation is only preliminar: in particular, only a subset of electrogeodesics has been discussed, namely those with $\ell=-\kappa \cP$. It reveals a rich phenomenology, which needs to be studied in detail. A full classification of the different possible trajectories of charged particles will be discussed elsewhere. It would also be interesting to confront the present findings with numerical GR/PIC simulations of this magnetosphere. Indeed, the present model can serve as a background to perform a perturbative study of the magnetosphere and the ejection of particles. Finally, it would be interesting to explore how the present approach can be extended to more general magnetized rotating black hole solutions of general relativity that still exhibit Killing-Yano symmetries. See for instance \cite{Ovcharenko:2025cpm}. We are leaving these different directions for future works. 
\newpage
\section*{Acknowledgments}

J. BA is grateful to the University of Santiago de Chile (USACH) for its hospitality during his different visits where this project was worked out. J.BA also warmly thank L\'ea Jouvin for enjoyable and inspiring discussions on the galactic center and the $\gamma$-astronomy on the road to Tilcara. J. BA and I. EM are also grateful to Abdelkader Abdallah for motivating discussions in the first step of this work. I.EM acknowledges support from the grant ANID FONDECYT 11240206 and funding via the BASAL Centro de Excelencia en Astrofisica y Tecnologias Afines (CATA) grant PFB06/2007.
E.G. acknowledges funding by l’Agence Nationale de la Recherche, projects StronG ANR-22-CE31-0015-01 and Einstein PPF ANR-23-CE40-0010-02.

\appendix

\section{SageMath notebooks} \label{sage_notebooks}

Some computations performed in this article have been performed by means of the free Python-based mathematical software system SageMath \cite{SageMath,SageManifolds}. 
The relevant Jupyter notebooks are publicly available from the following links: 

\begin{enumerate}
    \item Killing and Killing-Yano magnetospheres around a Kerr black hole:\\
    \url{https://nbviewer.org/url/relativite.obspm.fr/notebooks/KY_magnetosphere.ipynb}
    \item The Kerr monopole magnetosphere and its electrogeodesics:\\
    \url{https://nbviewer.org/url/relativite.obspm.fr/notebooks/Kerr_monopole.ipynb}
\end{enumerate}


\begin{thebibliography}{ab}


\bibitem{Doeleman:2012zc}
S.~S.~Doeleman, V.~L.~Fish, D.~E.~Schenck, C.~Beaudoin, R.~Blundell, G.~C.~Bower, A.~E.~Broderick, R.~Chamberlin, R.~Freund and P.~Friberg, \textit{et al.}
``Jet Launching Structure Resolved Near the Supermassive Black Hole in M87,''
Science \textbf{338}, 355 (2012)
doi:10.1126/science.1224768
[arXiv:1210.6132 [astro-ph.HE]].


\bibitem{Hada:2024icg}
K.~Hada, K.~Asada, M.~Nakamura and M.~Kino,
``M~87: a cosmic laboratory for deciphering black hole accretion and jet formation,''
Astron. Astrophys. Rev. \textbf{32}, no.1, 5 (2024)
doi:10.1007/s00159-024-00155-y
[arXiv:2412.07083 [astro-ph.HE]].

\bibitem{Oei:2024prz}
M.~S.~S.~L.~Oei, M.~J.~Hardcastle, R.~Timmerman, A.~R.~D.~J.~G.~I.~B.~Gast, A.~Botteon, A.~C.~Rodriguez, D.~Stern, G.~Calistro Rivera, R.~J.~van Weeren and H.~J.~A.~R{\"o}ttgering, \textit{et al.}
``Black hole jets on the scale of the cosmic web,''
Nature \textbf{633}, no.8030, 537-541 (2024)
doi:10.1038/s41586-024-07879-y
[arXiv:2411.08630 [astro-ph.HE]].


\bibitem{Blandford:2018iot}
R.~Blandford, D.~Meier and A.~Readhead,
``Relativistic Jets from Active Galactic Nuclei,''
Ann. Rev. Astron. Astrophys. \textbf{57}, 467-509 (2019)
doi:10.1146/annurev-astro-081817-051948
[arXiv:1812.06025 [astro-ph.HE]].



\bibitem{Lister:2016ojc}
M.~L.~Lister, M.~F.~Aller, H.~D.~Aller, D.~C.~Homan, K.~I.~Kellermann, Y.~Y.~Kovalev, A.~B.~Pushkarev, J.~L.~Richards, E.~Ros and T.~Savolainen,
``MOJAVE XIII. Parsec-Scale AGN Jet Kinematics Analysis Based on 19 years of VLBA Observations at 15 GHz,''
Astron. J. \textbf{152}, 12 (2016)
doi:10.3847/0004-6256/152/1/12
[arXiv:1603.03882 [astro-ph.GA]].

\bibitem{Lister:2021wmw}
M.~L.~Lister, D.~C.~Homan, K.~I.~Kellermann, Y.~Y.~Kovalev, A.~B.~Pushkarev, E.~Ros and T.~Savolainen,
``Monitoring Of Jets in Active Galactic Nuclei with VLBA Experiments. XVIII. Kinematics and Inner Jet Evolution of Bright Radio-loud Active Galaxies,''
Astrophys. J. \textbf{923}, no.1, 30 (2021)
[erratum: Astrophys. J. \textbf{949}, no.1, 34 (2023)]
doi:10.3847/1538-4357/ac230f
[arXiv:2108.13358 [astro-ph.HE]].

\bibitem{EventHorizonTelescope:2019dse}
K.~Akiyama \textit{et al.} [Event Horizon Telescope],
``First M87 Event Horizon Telescope Results. I. The Shadow of the Supermassive Black Hole,''
Astrophys. J. Lett. \textbf{875}, L1 (2019)
doi:10.3847/2041-8213/ab0ec7
[arXiv:1906.11238 [astro-ph.GA]].

\bibitem{EventHorizonTelescope:2022wok}
K.~Akiyama \textit{et al.} [Event Horizon Telescope],
``First Sagittarius A* Event Horizon Telescope Results. III. Imaging of the Galactic Center Supermassive Black Hole,''
Astrophys. J. Lett. \textbf{930}, no.2, L14 (2022)
doi:10.3847/2041-8213/ac6429
[arXiv:2311.09479 [astro-ph.HE]].


\bibitem{EventHorizonTelescope:2025uqi}
Saurabh \textit{et al.} [Event Horizon Telescope],
``Probing jet base emission of M87* with the 2021 Event Horizon Telescope observations,''
[arXiv:2512.08970 [astro-ph.HE]].

\bibitem{Nokhrina:2019sxv}
E.~E.~Nokhrina, L.~I.~Gurvits, V.~S.~Beskin, M.~Nakamura, K.~Asada and K.~Hada,
``M87 black hole mass and spin estimate through the position of the jet boundary shape break,''
Mon. Not. Roy. Astron. Soc. \textbf{489}, no.1, 1197-1205 (2019)
doi:10.1093/mnras/stz2116
[arXiv:1904.05665 [astro-ph.HE]].


\bibitem{Blandford:1977ds}
R.~D.~Blandford and R.~L.~Znajek,
``Electromagnetic extractions of energy from Kerr black holes,''
Mon. Not. Roy. Astron. Soc. \textbf{179}, 433-456 (1977)
doi:10.1093/mnras/179.3.433

\bibitem{Komissarov:2008yh}
S.~S.~Komissarov,
``Blandford-Znajek mechanism versus Penrose process,''
J. Korean Phys. Soc. \textbf{54}, 2503-2512 (2009)
doi:10.3938/jkps.54.2503
[arXiv:0804.1912 [astro-ph]].

\bibitem{Ruiz:2012te}
M.~Ruiz, C.~Palenzuela, F.~Galeazzi and C.~Bona,
``The Role of the ergosphere in the Blandford-Znajek process,''
Mon. Not. Roy. Astron. Soc. \textbf{423}, 1300-1308 (2012)
doi:10.1111/j.1365-2966.2012.20950.x
[arXiv:1203.4125 [gr-qc]].

\bibitem{Toma:2014kva}
K.~Toma and F.~Takahara,
``Electromotive force in the Blandford{\textendash}Znajek process,''
Mon. Not. Roy. Astron. Soc. \textbf{442}, no.4, 2855-2866 (2014)
doi:10.1093/mnras/stu1053
[arXiv:1405.7437 [astro-ph.HE]].

\bibitem{Kinoshita:2017mio}
S.~Kinoshita and T.~Igata,
``The essence of the Blandford{\textendash}Znajek process,''
PTEP \textbf{2018}, no.3, 033E02 (2018)
doi:10.1093/ptep/pty024
[arXiv:1710.09152 [gr-qc]].

\bibitem{Okamoto:2024fxa}
I.~Okamoto, T.~Uchida and Y.~Song,
``Electromagnetic Energy Extraction in Kerr Black Holes through Frame-Dragging Magnetospheres,''
[arXiv:2401.12684 [astro-ph.GA]].

\bibitem{Toma:2024tan}
K.~Toma, F.~Takahara and M.~Nakamura,
``On the Mechanism of Black Hole Energy Reduction in the Blandford{\textendash}Znajek Process,''
PTEP \textbf{2025}, no.3, 033E02 (2025)
doi:10.1093/ptep/ptaf036
[arXiv:2408.09993 [astro-ph.HE]].


\bibitem{Punsly:1989zz}
B.~Punsly and F.~V.~Coroniti,
``Electrodynamics of the event horizon,''
Phys. Rev. D \textbf{40}, 3834-3857 (1989)
doi:10.1103/PhysRevD.40.3834


\bibitem{Koide:1999bj}
S.~Koide, D.~L.~Meier, K.~Shibata and T.~Kudoh,
``General relativistic simulations of jet formation in a rapidly rotating black hole magnetosphere,''
Astrophys. J. \textbf{536}, 668 (2000)
doi:10.1086/308986
[arXiv:astro-ph/9907435 [astro-ph]].

\bibitem{Nishikawa:2004wp}
K.~I.~Nishikawa, G.~Richardson, S.~Koide, K.~Shibata, T.~Kudoh, P.~Hardee and G.~J.~Fishman,
``A General relativistic magnetohydrodynamics simulation of jet formation with a state transition,''
Astrophys. J. \textbf{625}, 60 (2005)
doi:10.1086/429360
[arXiv:astro-ph/0403032 [astro-ph]].

\bibitem{McKinney:2004ka}
J.~C.~McKinney and C.~F.~Gammie,
``A Measurement of the electromagnetic luminosity of a Kerr black hole,''
Astrophys. J. \textbf{611}, 977-995 (2004)
doi:10.1086/422244
[arXiv:astro-ph/0404512 [astro-ph]].

\bibitem{Komissarov:2004qu}
S.~S.~Komissarov,
``General relativistic MHD simulations of monopole magnetospheres of black holes,''
Mon. Not. Roy. Astron. Soc. \textbf{350}, 1431 (2004)
doi:10.1111/j.1365-2966.2004.07738.x
[arXiv:astro-ph/0402430 [astro-ph]].

\bibitem{Komissarov:2005wj}
S.~S.~Komissarov,
``Observations of the Blandford-Znajek and the MHD Penrose processes in computer simulations of black hole magnetospheres,''
Mon. Not. Roy. Astron. Soc. \textbf{359}, 801-808 (2005)
doi:10.1111/j.1365-2966.2005.08974.x
[arXiv:astro-ph/0501599 [astro-ph]].

\bibitem{Hawley:2005xs}
J.~F.~Hawley and J.~H.~Krolik,
``Magnetically driven jets in the kerr metric,''
Astrophys. J. \textbf{641}, 103-116 (2006)
doi:10.1086/500385
[arXiv:astro-ph/0512227 [astro-ph]].

\bibitem{Tchekhovskoy:2011zx}
A.~Tchekhovskoy, R.~Narayan and J.~C.~McKinney,
``Efficient Generation of Jets from Magnetically Arrested Accretion on a Rapidly Spinning Black Hole,''
Mon. Not. Roy. Astron. Soc. \textbf{418}, L79-L83 (2011)
doi:10.1111/j.1745-3933.2011.01147.x
[arXiv:1108.0412 [astro-ph.HE]].

\bibitem{Gralla:2014yja}
S.~E.~Gralla and T.~Jacobson,
``Spacetime approach to force-free magnetospheres,''
Mon. Not. Roy. Astron. Soc. \textbf{445}, no.3, 2500-2534 (2014)
[erratum: Mon. Not. Roy. Astron. Soc. \textbf{534}, no.2, 1541 (2024)]
doi:10.1093/mnras/stu1690
[arXiv:1401.6159 [astro-ph.HE]].

\bibitem{Camilloni:2022kmx}
F.~Camilloni, O.~J.~C.~Dias, G.~Grignani, T.~Harmark, R.~Oliveri, M.~Orselli, A.~Placidi and J.~E.~Santos,
``Blandford-Znajek monopole expansion revisited: novel non-analytic contributions to the power emission,''
JCAP \textbf{07}, no.07, 032 (2022)
doi:10.1088/1475-7516/2022/07/032
[arXiv:2201.11068 [gr-qc]].

\bibitem{Grignani:2018ntq}
G.~Grignani, T.~Harmark and M.~Orselli,
``Existence of the Blandford-Znajek monopole for a slowly rotating Kerr black hole,''
Phys. Rev. D \textbf{98}, no.8, 084056 (2018)
doi:10.1103/PhysRevD.98.084056
[arXiv:1804.05846 [gr-qc]].

\bibitem{Pan:2015haa}
Z.~Pan and C.~Yu,
``Fourth-order split monopole perturbation solutions to the Blandford-Znajek mechanism,''
Phys. Rev. D \textbf{91}, no.6, 064067 (2015)
doi:10.1103/PhysRevD.91.064067
[arXiv:1503.05248 [astro-ph.HE]].

\bibitem{Zhang:2014pla}
F.~Zhang, H.~Yang and L.~Lehner,
``Towards an understanding of the force-free magnetosphere of rapidly spinning black holes,''
Phys. Rev. D \textbf{90}, no.12, 124009 (2014)
doi:10.1103/PhysRevD.90.124009
[arXiv:1409.0345 [astro-ph.HE]].



\bibitem{Daniel:1997nu}
J.~Daniel and T.~Tajima,
``Electromagnetic waves in a strong Schwarzschild plasma,''
Phys. Rev. D \textbf{55}, 5193-5204 (1997)
doi:10.1103/PhysRevD.55.5193

\bibitem{Levinson:2018arx}
A.~Levinson and B.~Cerutti,
``Particle-in-cell simulations of pair discharges in a starved magnetosphere of a Kerr black hole,''
Astron. Astrophys. \textbf{616}, A184 (2018)
doi:10.1051/0004-6361/201832915
[arXiv:1803.04427 [astro-ph.HE]].

\bibitem{Chen:2018khs}
A.~Y.~Chen, Y.~Yuan and H.~Yang,
``Physics of Pair Producing Gaps in Black Hole Magnetospheres,''
Astrophys. J. Lett. \textbf{863}, no.2, L31 (2018)
doi:10.3847/2041-8213/aad8ab
[arXiv:1805.11039 [astro-ph.HE]].





\bibitem{Hirotani:2020ubl}
K.~Hirotani, R.~Krasnopolsky, H.~Shang, K.~i.~Nishikawa and M.~Watson,
``Two-dimensional Particle-in-Cell simulations of axisymmetric black hole magnetospheres,''
Astrophys. J. \textbf{908}, no.1, 88 (2021)
doi:10.3847/1538-4357/abd3a6
[arXiv:2012.07229 [astro-ph.HE]].



\bibitem{Yuan:2025war}
Y.~Yuan, A.~Y.~Chen and M.~Luepker,
``Physics of Pair-producing Gaps in Black Hole Magnetospheres: Two-dimensional General Relativistic Particle-in-cell Simulations,''
Astrophys. J. \textbf{985}, no.2, 159 (2025)
doi:10.3847/1538-4357/adce79
[arXiv:2503.08487 [astro-ph.HE]].

\bibitem{Chernoglazov:2023ksi}
A.~Chernoglazov, H.~Hakobyan and A.~A.~Philippov,
``High-energy Radiation and Ion Acceleration in Three-dimensional Relativistic Magnetic Reconnection with Strong Synchrotron Cooling,''
Astrophys. J. \textbf{959}, no.2, 122 (2023)
doi:10.3847/1538-4357/acffc6
[arXiv:2305.02348 [astro-ph.HE]].

\bibitem{Luepker:2025efw}
M.~Luepker, Y.~Yuan and A.~Y.~Chen,
``Stable Collisionless Tori around Kerr Black Holes,''
Astrophys. J. \textbf{996}, no.1, 75 (2026)
doi:10.3847/1538-4357/ae2326
[arXiv:2508.09077 [astro-ph.HE]].

\bibitem{Meringolo:2025bdu}
C.~Meringolo, F.~Camilloni and L.~Rezzolla,
``Electromagnetic Energy Extraction from Kerr Black Holes: Ab Initio Calculations,''
Astrophys. J. Lett. \textbf{992}, no.1, L8 (2025)
doi:10.3847/2041-8213/ae06a6
[arXiv:2507.08942 [gr-qc]].


\bibitem{Galishnikova:2025pfi}
A.~Galishnikova, H.~Hakobyan, A.~Philippov and B.~Crinquand,
``Entity{\textemdash}Hardware-agnostic Particle-in-cell Code for Plasma Astrophysics. II. General Relativistic Module,''
Astrophys. J. Supp. \textbf{285}, no.1, 12 (2026)
doi:10.3847/1538-4365/ae6592
[arXiv:2511.17701 [astro-ph.HE]].

\bibitem{Chen:2025qio}
A.~Y.~Chen, M.~Luepker and Y.~Yuan,
``Introducing APERTURE: A GPU-based General Relativistic Particle-in-Cell Simulation Framework,''
[arXiv:2503.04558 [astro-ph.HE]].

\bibitem{Meringolo:2026qjf}
C.~Meringolo and L.~Rezzolla,
``FPIC: A new Particle-In-Cell code for stationary and axisymmetric black-hole spacetimes,''
Astron. Comput. \textbf{57}, 101154 (2026)
doi:10.1016/j.ascom.2026.101154
[arXiv:2602.07452 [astro-ph.HE]].


\bibitem{Parfrey:2018dnc}
K.~Parfrey, A.~Philippov and B.~Cerutti,
``First-Principles Plasma Simulations of Black-Hole Jet Launching,''
Phys. Rev. Lett. \textbf{122}, no.3, 035101 (2019)
doi:10.1103/PhysRevLett.122.035101
[arXiv:1810.03613 [astro-ph.HE]].

\bibitem{Chen:2019osy}
A.~Y.~Chen and Y.~Yuan,
``Physics of Pair Producing Gaps in Black Hole Magnetospheres II -- General Relativity,''
Astrophys. J. \textbf{895}, no.2, 121 (2020)
doi:10.3847/1538-4357/ab8c46
[arXiv:1908.06919 [astro-ph.HE]].

\bibitem{Crinquand:2020ppq}
B.~Crinquand, B.~Cerutti, A.~Philippov, K.~Parfrey and G.~Dubus,
``Multidimensional Simulations of Ergospheric Pair Discharges around Black Holes,''
Phys. Rev. Lett. \textbf{124}, no.14, 145101 (2020)
doi:10.1103/PhysRevLett.124.145101
[arXiv:2003.03548 [astro-ph.HE]].


\bibitem{Crinquand:2020reu}
B.~Crinquand, B.~Cerutti, G.~Dubus, K.~Parfrey and A.~Philippov,
``Synthetic gamma-ray light curves of Kerr black hole magnetospheric activity from particle-in-cell simulations,''
Astron. Astrophys. \textbf{650}, A163 (2021)
doi:10.1051/0004-6361/202040158
[arXiv:2012.09733 [astro-ph.HE]].

\bibitem{ElMellah:2021tjo}
I.~El Mellah, B.~Cerutti, B.~Crinquand and K.~Parfrey,
``Spinning black holes magnetically connected to a Keplerian disk - Magnetosphere, reconnection sheet, particle acceleration, and coronal heating,''
Astron. Astrophys. \textbf{663}, A169 (2022)
doi:10.1051/0004-6361/202142847
[arXiv:2112.03933 [astro-ph.HE]].

\bibitem{ElMellah:2023sun}
I.~El Mellah, B.~Cerutti and B.~Crinquand,
``Reconnection-driven flares in 3D black hole magnetospheres - A scenario for hot spots around Sagittarius A*,''
Astron. Astrophys. \textbf{677}, A67 (2023)
doi:10.1051/0004-6361/202346781
[arXiv:2305.01689 [astro-ph.HE]].

\bibitem{Mehlhaff:2025mxi}
J.~Mehlhaff, B.~Cerutti and B.~Crinquand,
``A kinetic model of jet-corona coupling in accreting black holes,''
Astron. Astrophys. \textbf{701}, A62 (2025)
doi:10.1051/0004-6361/202453561
[arXiv:2504.01062 [astro-ph.HE]].

\bibitem{Figueiredo:2025xbo}
E.~Figueiredo, B.~Cerutti and K.~Parfrey,
``Effect of magnetic field inclination on black hole jet power and particle acceleration,''
Astron. Astrophys. \textbf{700}, L19 (2025)
doi:10.1051/0004-6361/202555826
[arXiv:2508.02211 [astro-ph.HE]].







\bibitem{deFelice:1999ch}
F.~de Felice and O.~Zanotti,
``Jet dynamics in black hole physics: Acceleration during subparsec collimation,''
Gen. Rel. Grav. \textbf{32}, 1449-1472 (2000)
doi:10.1023/A:1001925916715
[arXiv:astro-ph/9912413 [astro-ph]].



\bibitem{Gralla:2015wva}
S.~E.~Gralla and T.~Jacobson,
``Nonaxisymmetric Poynting Jets,''
Phys. Rev. D \textbf{92}, no.4, 043002 (2015)
doi:10.1103/PhysRevD.92.043002
[arXiv:1503.03848 [astro-ph.HE]].

\bibitem{Mizuno:2025mog}
M.~Mizuno, S.~E.~Gralla and A.~Philippov,
``Plasma flow in force-free magnetospheres: two-fluid model near pulsars and black holes,''
[arXiv:2511.02057 [astro-ph.HE]].

\bibitem{Kolos:2015iva}
M.~Kolo{\v{s}}, Z.~Stuchl{\'\i}k and A.~Tursunov,
``Quasi-harmonic oscillatory motion of charged particles around a Schwarzschild black hole immersed in a uniform magnetic field,''
Class. Quant. Grav. \textbf{32}, no.16, 165009 (2015)
doi:10.1088/0264-9381/32/16/165009
[arXiv:1506.06799 [gr-qc]].

\bibitem{Tursunov:2016dss}
A.~Tursunov, Z.~Stuchl{\'\i}k and M.~Kolo{\v{s}},
``Circular orbits and related quasiharmonic oscillatory motion of charged particles around weakly magnetized rotating black holes,''
Phys. Rev. D \textbf{93}, no.8, 084012 (2016)
doi:10.1103/PhysRevD.93.084012
[arXiv:1603.07264 [gr-qc]].

\bibitem{Kopacek:2018lgy}
O.~Kop{\'a}{\v{c}}ek and V.~Karas,
``Near-horizon structure of escape zones of electrically charged particles around weakly magnetized rotating black hole,''
Astrophys. J. \textbf{853}, no.1, 53 (2018)
doi:10.3847/1538-4357/aaa45f
[arXiv:1801.01576 [astro-ph.HE]].

\bibitem{Kopacek:2020scv}
O.~Kop{\'a}{\v{c}}ek and V.~Karas,
``Near-horizon structure of escape zones of electrically charged particles around weakly magnetized rotating black hole. II. Acceleration and escape in the oblique magnetosphere,''
Astrophys. J. \textbf{900}, no.2, 119 (2020)
doi:10.3847/1538-4357/ababa8
[arXiv:2008.04630 [astro-ph.HE]].

\bibitem{Kopacek:2021lnq}
O.~Kopacek and V.~Karas,
``Magnetized black holes: The role of rotation, boost, and accretion in twisting the field lines and accelerating particles,''
doi:10.1142/9789811269776{\_}0333
[arXiv:2111.00493 [astro-ph.HE]].

\bibitem{Capitanio:2022epf}
F.~Capitanio, S.~Fabiani, A.~Gnarini, F.~Ursini, C.~Ferrigno, G.~Matt, J.~Poutanen, M.~Cocchi, R.~Mikusincova and R.~Farinelli, \textit{et al.}
``Polarization Properties of the Weakly Magnetized Neutron Star X-Ray Binary GS 1826{\textendash}238 in the High Soft State,''
Astrophys. J. \textbf{943}, no.2, 129 (2023)
doi:10.3847/1538-4357/acae88
[arXiv:2212.12472 [astro-ph.HE]].




\bibitem{Khan:2023ttq}
S.~U.~Khan and Z.~M.~Chen,
``Charged particle dynamics in black hole split monopole magnetosphere,''
Eur. Phys. J. C \textbf{83} (2023) no.8, 704
[erratum: Eur. Phys. J. C \textbf{83} (2023) no.8, 760]

\bibitem{Kolos:2023oii}
M.~Kolo{\v{s}}, M.~Shahzadi and A.~Tursunov,
``Charged particle dynamics in parabolic magnetosphere around Schwarzschild black hole,''
Eur. Phys. J. C \textbf{83}, no.4, 323 (2023)
doi:10.1140/epjc/s10052-023-11498-8
[arXiv:2304.13603 [gr-qc]].

\bibitem{Rueda:2024xeb}
J.~A.~Rueda and R.~Ruffini,
``Kerr black hole energy extraction, irreducible mass feedback, and the effect of captured particles charge,''
Eur. Phys. J. C \textbf{84}, no.11, 1166 (2024)
doi:10.1140/epjc/s10052-024-13459-1
[arXiv:2410.04776 [gr-qc]].

\bibitem{Cherubini:2025lnc}
C.~Cherubini, R.~Moradi, J.~A.~Rueda and R.~Ruffini,
``Vacuum breakdown around a Kerr black hole surrounded by a magnetic field,''
doi:10.1051/0004-6361/202556413
[arXiv:2510.23886 [gr-qc]].


\bibitem{Dyson:2023ujk}
C.~Dyson and D.~Pere{\~n}iguez,
``Magnetic black holes: From Thomson dipoles to the Penrose process and cosmic censorship,''
Phys. Rev. D \textbf{108} (2023) no.8, 084064
doi:10.1103/PhysRevD.108.084064
[arXiv:2306.15751 [gr-qc]].

\bibitem{Carter:1968rr}
B.~Carter,
``Global structure of the Kerr family of gravitational fields,''
Phys. Rev. \textbf{174}, 1559-1571 (1968)
doi:10.1103/PhysRev.174.1559

\bibitem{Hackmann:2013pva}
E.~Hackmann and H.~Xu,
``Charged particle motion in Kerr-Newman space-times,''
Phys. Rev. D \textbf{87}, no.12, 124030 (2013)
doi:10.1103/PhysRevD.87.124030
[arXiv:1304.2142 [gr-qc]].

\bibitem{Grunau:2010gd}
S.~Grunau and V.~Kagramanova,
``Geodesics of electrically and magnetically charged test particles in the Reissner-Nordstr{\"o}m space-time: analytical solutions,''
Phys. Rev. D \textbf{83}, 044009 (2011)
doi:10.1103/PhysRevD.83.044009
[arXiv:1011.5399 [gr-qc]].

\bibitem{Olivares:2011xb}
M.~Olivares, J.~Saavedra, J.~R.~Villanueva and C.~Leiva,
``Motion of charged particles on the Reissner-Nordstr{\"o}m (Anti)-de Sitter black holes,''
Mod. Phys. Lett. A \textbf{26}, 2923-2950 (2011)
doi:10.1142/S0217732311037261
[arXiv:1101.0748 [gr-qc]].

\bibitem{Das:2016opi}
P.~Das, R.~Sk and S.~Ghosh,
``Motion of charged particle in Reissner{\textendash}Nordstr{\"o}m spacetime: a Jacobi-metric approach,''
Eur. Phys. J. C \textbf{77}, no.11, 735 (2017)
doi:10.1140/epjc/s10052-017-5295-6
[arXiv:1609.04577 [gr-qc]].


\bibitem{Pugliese:2013zma}
D.~Pugliese, H.~Quevedo and R.~Ruffini,
``Equatorial circular orbits of neutral test particles in the Kerr-Newman spacetime,''
Phys. Rev. D \textbf{88}, no.2, 024042 (2013)
doi:10.1103/PhysRevD.88.024042
[arXiv:1303.6250 [gr-qc]].

\bibitem{Cui:2023uyb}
Y.~Cui, K.~Hada, T.~Kawashima, M.~Kino, W.~Lin, Y.~Mizuno, H.~Ro, M.~Honma, K.~Yi and J.~Yu, \textit{et al.}
``Precessing jet nozzle connecting to a spinning black hole in M87,''
Nature \textbf{621} (2023), 711-715
doi:10.1038/s41586-023-06479-6
[arXiv:2310.09015 [astro-ph.HE]].

\bibitem{Tsagas:2004kv}
C.~G.~Tsagas,
``Electromagnetic fields in curved spacetimes,''
Class. Quant. Grav. \textbf{22}, 393-408 (2005)
doi:10.1088/0264-9381/22/2/011
[arXiv:gr-qc/0407080 [gr-qc]]

\bibitem{Krtous:2007xg}
P.~Krtous,
``Electromagnetic field in higher-dimensional black-hole spacetimes,''
Phys. Rev. D \textbf{76}, 084035 (2007)
doi:10.1103/PhysRevD.76.084035
[arXiv:0707.0002 [hep-th]].

\bibitem{Jezierski:2005cg}
J.~Jezierski and M.~Lukasik,
``Conformal Yano-Killing tensor for the Kerr metric and conserved quantities,''
Class. Quant. Grav. \textbf{23}, 2895-2918 (2006)
doi:10.1088/0264-9381/23/9/008
[arXiv:gr-qc/0510058 [gr-qc]].

\bibitem{Penrose:1982wp}
R.~Penrose,
``Quasilocal mass and angular momentum in general relativity,''
Proc. Roy. Soc. Lond. A \textbf{381}, 53-63 (1982)
doi:10.1098/rspa.1982.0058


\bibitem{Jezierski:2015lwa}
J.~Jezierski and T.~Smo\l{}ka,
``A geometric description of Maxwell field in a Kerr spacetime,''
Class. Quant. Grav. \textbf{33}, no.12, 125035 (2016)
doi:10.1088/0264-9381/33/12/125035
[arXiv:1502.00599 [gr-qc]].

\bibitem{Kolar:2015cha}
I.~Kolar and P.~Krtous,
``Weak electromagnetic field admitting integrability in Kerr-NUT-(A)dS spacetimes,''
Phys. Rev. D \textbf{91}, no.12, 124045 (2015)
doi:10.1103/PhysRevD.91.124045
[arXiv:1504.00524 [gr-qc]].



\bibitem{Wald:1974np}
R.~M.~Wald,
``Black hole in a uniform magnetic field,''
Phys. Rev. D \textbf{10}, 1680-1685 (1974)
doi:10.1103/PhysRevD.10.1680

\bibitem{Michel}
F.C.~Michel, ''Rotating Magnetosphere: a Simple Relativistic Model'', ApJ \textbf{180}, 207 (1973).





\bibitem{Bicak:2006hs}
J.~Bi{\v{c}}{\'a}k, V.~Karas and T.~Ledvinka,
``Black holes and magnetic fields,''
IAU Symp. \textbf{238}, 139-144 (2007)
doi:10.1017/S1743921307004851
[arXiv:astro-ph/0610841 [astro-ph]].





\bibitem{Krtous:2006qy}
P.~Krtous, D.~Kubiznak, D.~N.~Page and V.~P.~Frolov,
``Killing-Yano Tensors, Rank-2 Killing Tensors, and Conserved Quantities in Higher Dimensions,''
JHEP \textbf{02}, 004 (2007)
doi:10.1088/1126-6708/2007/02/004
[arXiv:hep-th/0612029 [hep-th]].

\bibitem{Kubiznak:2006kt}
D.~Kubiznak and V.~P.~Frolov,
``Hidden Symmetry of Higher Dimensional Kerr-NUT-AdS Spacetimes,''
Class. Quant. Grav. \textbf{24}, no.3, F1-F6 (2007)
doi:10.1088/0264-9381/24/3/F01
[arXiv:gr-qc/0610144 [gr-qc]].

\bibitem{Krtous:2008tb}
P.~Krtous, V.~P.~Frolov and D.~Kubiznak,
``Hidden Symmetries of Higher Dimensional Black Holes and Uniqueness of the Kerr-NUT-(A)dS spacetime,''
Phys. Rev. D \textbf{78}, 064022 (2008)
doi:10.1103/PhysRevD.78.064022
[arXiv:0804.4705 [hep-th]].


\bibitem{Frolov:2017kze}
V.~P.~Frolov, P.~Krtous and D.~Kubiznak,
``Black holes, hidden symmetries, and complete integrability,''
Living Rev. Rel. \textbf{20}, no.1, 6 (2017)
doi:10.1007/s41114-017-0009-9
[arXiv:1705.05482 [gr-qc]].



\bibitem{Carter:1987id}
B.~Carter,
``Separability of the Killing Maxwell System Underlying the Generalized Angular Momentum Constant in the {Kerr-Newman} Black Hole Metrics,''
J. Math. Phys. \textbf{28}, 1535-1538 (1987)
doi:10.1063/1.527509





















\bibitem{Visinescu:2009rm}
M.~Visinescu,
``Higher order first integrals of motion in a gauge covariant Hamiltonian framework,''
Mod. Phys. Lett. A \textbf{25}, 341-350 (2010)
doi:10.1142/S0217732310032500
[arXiv:0910.3474 [hep-th]].

\bibitem{Igata:2010ny}
T.~Igata, T.~Koike and H.~Ishihara,
``Constants of Motion for Constrained Hamiltonian Systems: A Particle around a Charged Rotating Black Hole,''
Phys. Rev. D \textbf{83}, 065027 (2011)
doi:10.1103/PhysRevD.83.065027
[arXiv:1005.1815 [gr-qc]].

\bibitem{Igata:2010bu}
T.~Igata, T.~Koike and H.~Ishihara,
``Killing Tensors and Conserved Quantities of a Relativistic Particle in External Fields,''
doi:10.1142/9789814374552{\_}0121
[arXiv:1003.0791 [gr-qc]].


\bibitem{Obukhov:2020sjc}
V.~Obukhov,
``Hamilton{\textendash}Jacobi Equation for a Charged Test Particle in the St{\"a}ckel Space of Type (2.0),''
Symmetry \textbf{12}, no.8, 1289 (2020)
doi:10.3390/sym12081289
[arXiv:2007.09492 [gr-qc]].


\bibitem{Obukhov:2021ggp}
V.~V.~Obukhov,
``Algebras of integrals of motion for the Hamilton{\textendash}Jacobi and Klein{\textendash}Gordon{\textendash}Fock equations in spacetime with four-parameter groups of motions in the presence of an external electromagnetic field,''
J. Math. Phys. \textbf{63}, no.2, 023505 (2022)
doi:10.1063/5.0080703
[arXiv:2112.15138 [math-ph]].

\bibitem{Obukhov:2023uxi}
V.~V.~Obukhov,
``Hamilton-Jacobi and Klein-Gordon-Fock equations for a charged test particle in space-time with simply transitive four-parameter groups of motions,''
J. Math. Phys. \textbf{64}, no.9, 093507 (2023)
doi:10.1063/5.0158054
[arXiv:2309.14682 [math-ph]].

\bibitem{Ovcharenko:2025cpm}
H.~Ovcharenko and J.~Podolsk{\'y},
``New class of rotating charged black holes with nonaligned electromagnetic field,''
Phys. Rev. D \textbf{112} (2025) no.6, 064076
doi:10.1103/8wkz-th6v
[arXiv:2508.04850 [gr-qc]].









\bibitem{SageMath}
The SageMath Developers, SageMath (2026), \url{https://www.sagemath.org}.

\bibitem{SageManifolds}
SageManifolds (2026), \url{https://sagemanifolds.obspm.fr/}.

\end{thebibliography}
\end{document}